\documentclass[useAMS,usenatbib]{mn2e}

\newcommand{\aap}{{\it A\&A}}
\newcommand{\apss}{{\it Ap\&SS}}
\newcommand{\nat}{{\it Nature}}
\newcommand{\apj}{{\it ApJ}}
\newcommand{\apjl}{{\it ApJL}}
\newcommand{\pss}{{\it P\&SS}}
\newcommand{\mnras}{{ \it MNRAS}}
\newcommand{\science}{{ \it Science}}
\newcommand{\icarus}{{\it Icarus}}
\newcommand{\kepler}{{\it Kepler}}

\usepackage{graphicx}
\usepackage{epsfig}
\usepackage{amsmath}
\usepackage{color}
\usepackage{natbib}
\usepackage{comment}
\usepackage{txfonts}
\usepackage[update,prepend]{epstopdf}
\usepackage{subcaption}
\usepackage[pdftex,pdfpagemode={UseOutlines},bookmarks,bookmarksopen,colorlinks,linkcolor={blue},citecolor={blue},urlcolor={red}]{hyperref}

\title[The dearth of planets around the tightest binaries]{No circumbinary planets transiting the tightest Kepler binaries --- a possible fingerprint of a third star}

\author[Martin, Mazeh \& Fabrycky]
{\parbox{\textwidth}{David V. Martin$^1$\thanks{E-mail: david.martin@unige.ch}, Tsevi Mazeh$^{2}$ and Daniel C. Fabrycky$^{3}$}
\vspace{0.4cm}\\
\parbox{\textwidth}{$^{1}$Observatoire de Gen\`eve, Universit\'e de Gen\`eve, 51 chemin des Maillettes, Sauverny 1290, Switzerland\\
$^{2}$School of Physics and Astronomy, Raymond and Beverly Sackler Faculty of Exact Sciences, Tel Aviv University, Tel Aviv 69978, Israel\\
$^{3}$Department of Astronomy and Astrophysics, University of Chicago, 5640 South Ellis Avenue, Chicago, IL 60637, USA\\}}

\begin{document}
\date{Accepted . Received}

\pagerange{\pageref{firstpage}--\pageref{lastpage}} \pubyear{2015}

\maketitle

\label{firstpage}
\begin{abstract}
The \kepler\ mission has yielded the discovery of eight circumbinary systems, all found around eclipsing binaries with periods greater than 7 d. This is longer than the typical eclipsing binary period found by \kepler, and hence there is a dearth of planets around the closest binaries. In this paper we suggest how this dearth may be explained by the presence of a distant stellar tertiary companion, which shrunk the inner binary orbit by the process of Kozai cycles and tidal friction, a mechanism that has been implicated for producing most binaries with periods below 7 d.  We show that the geometry and orbital dynamics of these evolving triple-star systems are highly restrictive for a circumbinary planet, which is subject itself to Kozai modulation, on one hand, and can shield the two inner stars from their Kozai cycle and subsequent shrinking, on the other hand. Only small planets on wide and inclined orbits may form, survive and allow for the inner binary shrinkage. Those are difficult to detect.
\end{abstract}

\begin{keywords}
binaries: close, eclipsing -- astrometry and celestial mechanics: celestial mechanics -- planets and satellites: detection, dynamical evolution and stability -- methods: analytical, numerical
\end{keywords}

%===================
\section{Introduction}
\label{sec:intro}
%===================

% More recent reference than Cassan?

The first 
two decades of exoplanetary science have yielded many surprising results. Not only do most stars host orbiting planets \citep{cassan12}, but planets are often found in unexpected locations and with unexpected properties. 
For example, 
hot-Jupiters continue to pose significant theoretical challenges \citep{triaud10,madhusudhan14},
while Super-Earths were predicted not to form, and yet they are some of the most abundant planets known today \citep{howard10,mayor11}. Planets also have been found in binary star systems orbiting one (e.g., 16 Cygni, \citealt{cochran97}) and two (e.g., Kepler-16, \citealt{doyle11}) stars. The parameter space of non-discoveries is shrinking
fast. The field has therefore evolved to a state where an absence of planets is just as telling as a new discovery. 

One conspicuous absence is seen in the \kepler\ circumbinary planets (CBPs). So far there have been ten transiting CBPs discovered by \kepler\ orbiting eight eclipsing binaries (EBs), including the three-planet system Kepler-47 \citep{orosz12b}. 
It was pointed out by \citet{welsh14a} that all of the planets have been found orbiting EBs of periods between 7.4 and 40 d, despite the median of the EB catalog being 2.7 d. The discoveries have therefore been made on the tail of the EB period distribution. The analyses of \citet{armstrong14} and \citet{martintriaud14} suggest that this dearth of planets is statistically significant.

%---------------------------------------------------------------------
\begin{table*}
\caption{The \kepler\ transiting circumbinary planets}
\centering % centering table
\begin{tabular}{l | c | c | c | c | c | c | c | c | c | c | c | c }
\hline\hline %inserting double-line
Name        & $M_1$         & $M_2$       & $a_{\rm in}$ &
$P_{\rm in}$& $e_{\rm in}$ & $R_{\rm p}$ & $a_{\rm p}$& $P_{\rm p}$& $e_{\rm p}$ &
$ \Delta I_{\rm p,in}$ & $a_{\rm crit}$ & Reference            \\
 &[$M_\odot$] &[$M_\odot$]&[AU]                &[d]
&                         &$[R_\oplus$]&
[AU]               &[d]           &                      & [deg] &[AU]
&\\
%[0.5ex]
\hline % inserts single-line
16 & 0.69 & 0.20 & 0.22 & 40.1 & 0.16 & 8.27 & 0.71 & 228.8 & 0.01 &  0.31 & 0.64 & \citet{doyle11} \\
34 & 1.05 & 1.02 & 0.23 & 28.0 & 0.52 & 8.38 & 1.09 & 288.8 & 0.18 & 1.86 & 0.84 & \citet{welsh12} \\
35 & 0.89 & 0.81 & 0.18 & 20.7 & 0.14 & 7.99 & 0.60 & 131.4 & 0.04 & 1.07 & 0.50 & \citet{welsh12} \\
38 & 0.95 & 0.26 & 0.15 & 18.8 & 0.10 & 4.35 & 0.47 & 106.0 & 0.07 & 0.18 & 0.39 & \citet{orosz12a} \\
47b & 1.04 & 0.36 & 0.08 & 7.4 & 0.02 & 2.98 & 0.30 & 49.5 & 0.04 & 0.27 & 0.20 & \citet{orosz12b} \\
47d & 1.04 & 0.36 & 0.08 & 7.4 & 0.02 & --- & 0.72 & 187.3 & --- & --- & 0.20 & Orosz (in prep) \\
47c & 1.04 & 0.36 & 0.08 & 7.4 & 0.02 & 4.61 & 0.99 & 303.1 & $<$ 0.41 & 1.16 & 0.20 & \citet{orosz12b} \\
PH-1/64 & 1.50 & 0.40 & 0.18 & 20.0 & 0.21 & 6.18 & 0.65 & 138.5 & 0.07 & 2.81 & 0.54 & \citet{schwamb13,kostov13}\\
413 & 0.82 & 0.54 & 0.10 & 10.1 & 0.04 & 4.34 & 0.36 & 66.3 & 0.12 & 4.02 & 0.26 & \citet{kostov14}\\
3151 & 0.93 & 0.19 & 0.18 & 27.3 & 0.05 & 6.17 & 0.79 & 240.5 & 0.04 & 2.90 & 0.44 & \citet{welsh14b}\\
\hline % inserts single-line
\end{tabular}
\end{table*}
%--------------------------------------------------

In this paper we show how this absence of planets may be a natural consequence of close binary formation, formalising an explanation that was briefly proposed by \citet{welsh14a}, \citet{armstrong14}, \citet{martintriaud14} and \citet{winn14}.  A popular theory proposes that most very close binaries ($\lesssim7$ d) are initially formed at wider separations. The binaries subsequently shrink under the influence of a misaligned tertiary star and a process known as Kozai cycles with tidal friction (KCTF) (e.g., \citealt{mazeh79,fabrycky07}). The dynamics of such evolving triple star systems impose strict stability constraints on any putative circumbinary planetary orbits around the inner binary. We will demonstrate that these constraints act to either completely inhibit planetary formation and survival or restrict it to small planets on wide, misaligned orbits. 

We point out that \citet{munoz15} and \citet{hamers15b} have conducted independent analyses with complementary techniques, reaching the same general conclusions as this paper.

The plan of this paper is as follows: In Sect.~\ref{sec:sample} we discuss some of the relevant trends seen in the \kepler\ circumbinary planets. Sect.~\ref{sec:close_binary_formation} is a review of the theory and observations related to close binary formation via KCTF. In Sect.~\ref{sec:planetary_dynamics_multi-stellar} we summarise some important theoretical aspects of three-body dynamics and stability in binary star systems, and then extend the discussion to the triple star case. In Sect.~\ref{sec:example} we run n-body simulations on a set of example systems to test planet stability and binary shrinkage. We then discuss the protoplanetary disc environment and the effects of planet migration and multi-planetary systems in Sect.~\ref{sec:additional_effects}. In Sect.~\ref{sec:summary} we conclude by summarising the general argument and looking ahead to future complementary observations.

%====================
%Section 2
\section{Circumbinary planets discovered by \kepler}
\label{sec:sample}
%=====================

%----------------------------------------------------------------
% Figure 1

\begin{figure*}  
\begin{center}  
	\begin{subfigure}[b]{0.49\textwidth}
		\caption{}
		\includegraphics[width=\textwidth]{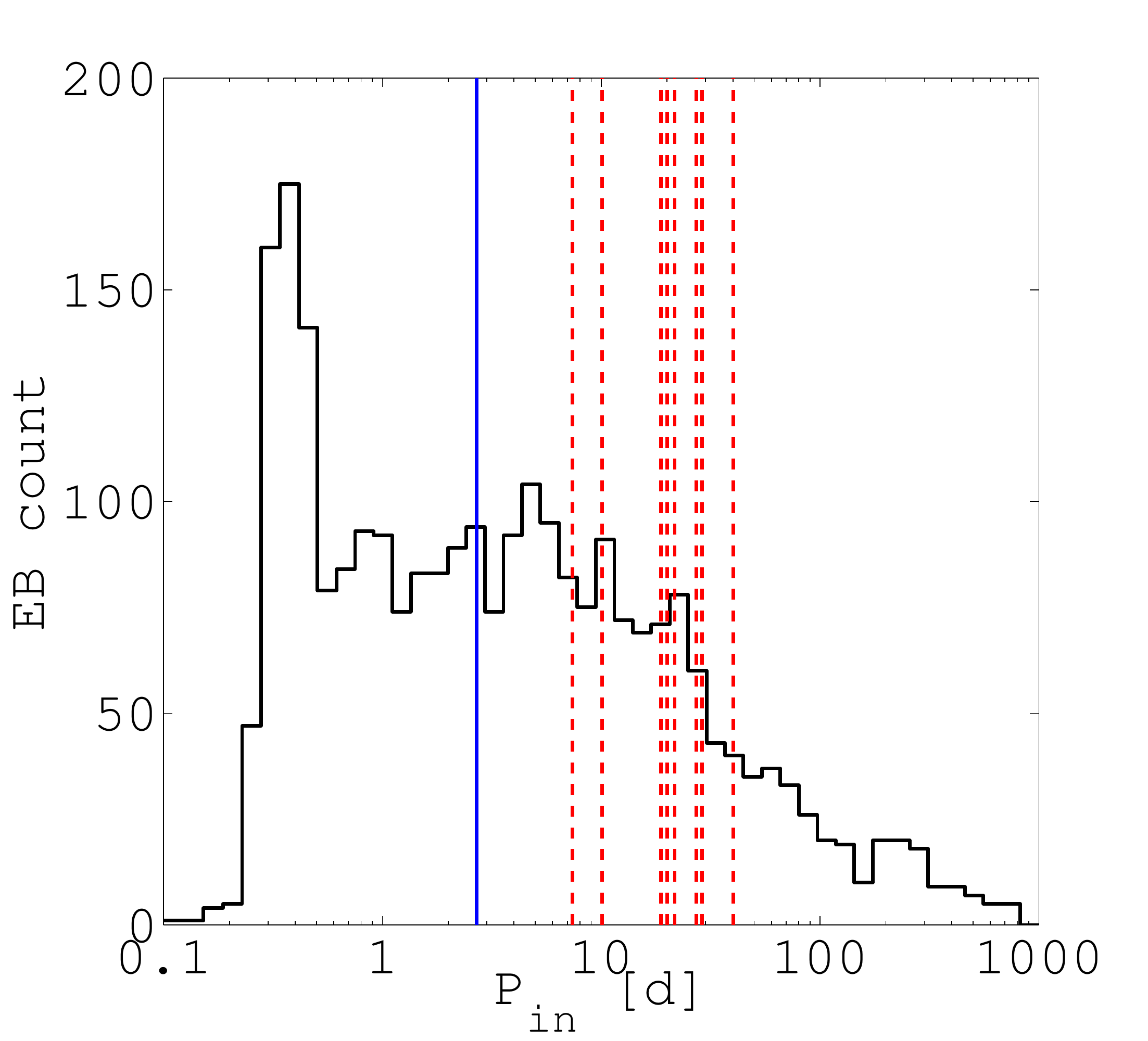}  
		\label{fig:binary_histogram}  
	\end{subfigure}
%	\begin{subfigure}[b]{0.33\textwidth}
%		\caption{}
%		\includegraphics[width=\textwidth]{Pbin_Pp-eps-converted-to.pdf}  
%		\label{fig:Pbin_Pp}  
%	\end{subfigure}
	\begin{subfigure}[b]{0.49\textwidth}
		\caption{}
		\includegraphics[width=\textwidth]{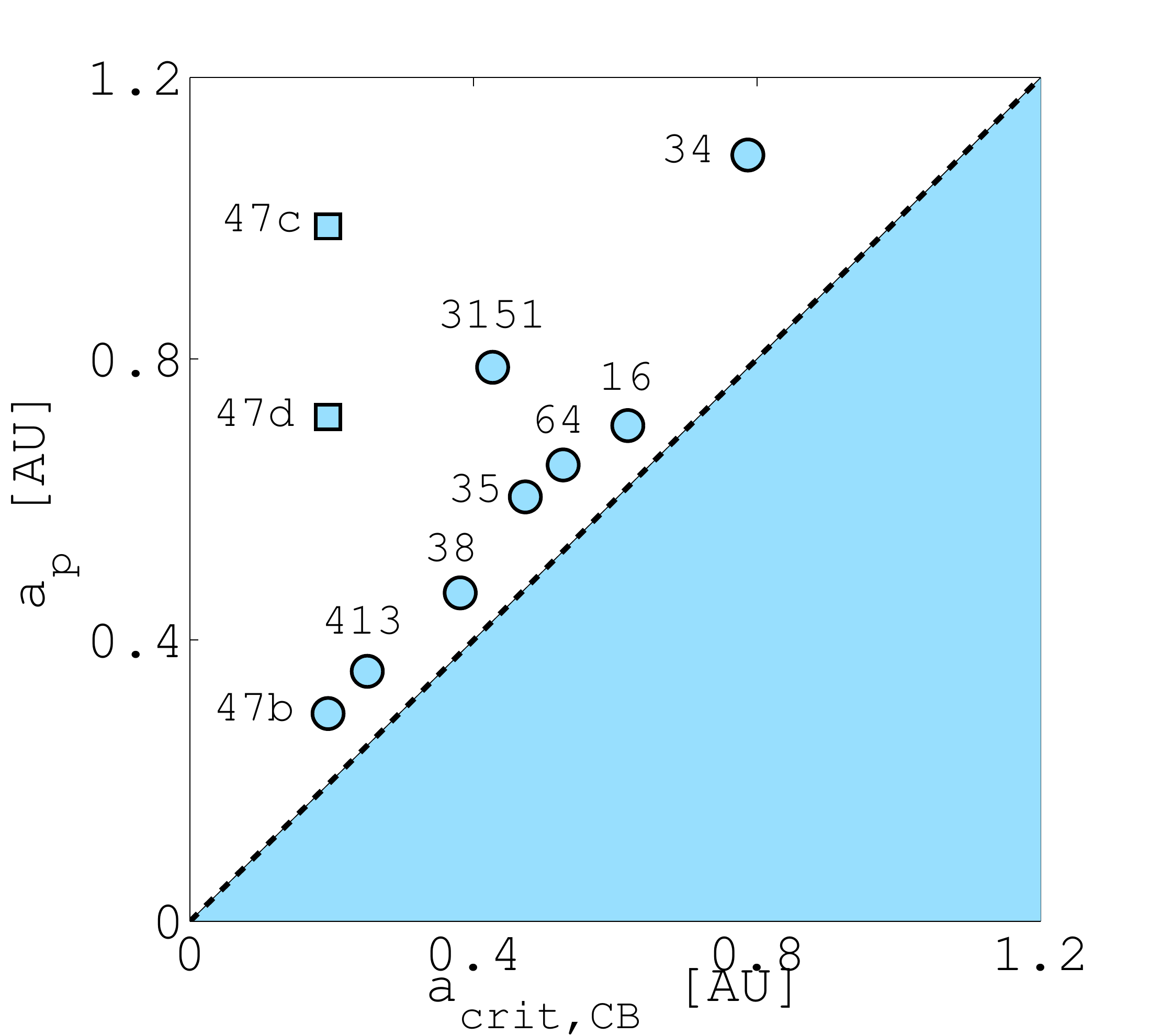}  
		\label{fig:stability_limit}  
	\end{subfigure}
	\caption{Circumbinary planets discovered so far via \kepler\ transits: 
(a) A histogram of periods of the \kepler\ EB catalog, 
with a blue solid vertical line indicating the median of this population.
Red dashed vertical lines indicating the binary periods known to host circumbinary planets, (b) Planet semi-major axis vs critical semi-major axis (Eq.~\ref{eq:stability_limit_CB}), where the blue shaded region corresponds to unstable orbits. Systems have been labeled by their Kepler number. The outer planets of  Kepler-47 are presented by square symbols.
}\label{fig:kepler_data}  
\end{center}  
\end{figure*} 
%----------------------------------------------------------------

The \kepler\ mission provided four years of near-continuous photometry of roughly 200,000 stars. As of May 8, 2015 there have been 2773 \kepler\ eclipsing binaries identified (\citealt{prsa11,slawson11}, Kirk et al. in prep)\footnote{The latest version of the catalog can be found online at http://keplerebs.villanova.edu, maintained by Prsa et al. at Villanova University.}. A search for additional transit signals has led to the discovery of eight circumbinary systems, one of which contains three planets (Kepler-47) (\citealt{welsh14a} and references there-in).

In Table~1 we summarise the parameters of the 10 transiting CBPs discovered by  \kepler. We list the masses of the two stellar components, $M_1$ and $M_2$, the semi-major axis, $a_{\rm in}$, period, $P_{\rm in}$, and eccentricity, $e_{\rm in}$, of the inner binary orbit. We then list the planetary parameters: its radius, $R_{\rm p}$, semi-major axis,  $a_{\rm p}$, period, $P_{\rm p}$, eccentricity, $e_{\rm p}$ and mutual inclination with respect to the inner binary orbital plane, $\Delta I_{\rm p,in}$. We then list the inner-most stable orbit for a circumbinary planet, $a_{\rm crit,CB}$, calculated according to \citet{holman99} (see Sect.~\ref{sec:planets_binaries_stability} for details). The last column is the discovery paper reference, including the two independent discoveries of PH-1/Kepler-64 by \citet{schwamb13}\footnote{The planet hunters consortium at http://www.planethunters.org/.} and \citet{kostov13}. The planet masses are not listed because they are generally poorly constrained.

In the small sample of CBPs some preliminary trends have been identified:

\begin{enumerate}
  \item The planets have all been discovered orbiting eclipsing binaries of relatively long periods. In Fig.~\ref{fig:binary_histogram} we demonstrate this feature by plotting a histogram of the  \kepler\ eclipsing binary periods, which has a median of 2.7 d (blue solid vertical line). The red dashed vertical lines indicate the binary periods around which planets have been found, of which the shortest is $P_{\rm in} =7.4$ d (Kepler-47). \citet{armstrong14} analysed the \kepler\ lightcurves using an automated search algorithm and debiasing process, and came to the conclusion that planets are significantly rarer around binaries of periods between 5 and 10 d than around wider binaries. \citet{martintriaud14} created synthetic circumbinary distributions and simulated eclipses and transits observable by \kepler. 
They concluded that the number of discoveries would have been roughly doubled if the circumbinary abundance did not drop around binaries with periods shorter than 5 d.

 \item Most of the circumbinary planets are found near the dynamical stability limit, which we demonstrate in Fig.~\ref{fig:stability_limit}. \citet{welsh14a} proposed that this was either the result of a physical pile-up of planets or detection biases, although \citet{martintriaud14} argued against the latter explanation.
\item The systems are all close to coplanarity. Out of the ten circumbinary planets found so far, the most misaligned is Kepler-413 with $\Delta I_{\rm p,in} = 4^{\circ}$ \citep{kostov14}. The mean misalignment is $1.7^{\circ}$. This is however an observational bias imposed by the requirement of consecutive transits in the detection method. Significantly misaligned systems ($\gtrsim 10^{\circ}$) would either avoid transiting or only do so irregularly \citep{martintriaud14}, hindering the ability to detect them.

 \item The planets have only been found within a narrow size range between 3 and 8.3 $R_{\oplus}$. The lack of Jupiter-size and larger planets might be due to the relative rarity of such big planets and the small number of detections. The paucity of small planets is probably a result of the large variations in circumbinary transit timing and duration, making it difficult to phase-fold the photometric data and identify shallow transits.

 \item Preliminary calculations indicate that the circumbinary abundance, within the present realm of detections, is $\sim 10\%$ \citep{armstrong14,martintriaud14}, which is similar to the 13\% abundance of Neptune-mass and heavier planets around single stars derived from radial velocity surveys \citep{mayor11}. This circumbinary abundance was calculated based on only the near-coplanar population that {\it Kepler} is sensitive to. If there exists a presently-hidden population of misaligned circumbinary planets then this would increase their overall abundance.

  %\item There are no known close stellar companions to any of the systems, although the non-detections are not very constraining. In the PH-1/Kepler-64 system there are two additional stars, themselves in a close binary, but at a distance of $\sim 1000$ AU from the circumbinary system \citep{schwamb13,kostov13}, which is probably too far away to affect the planet formation process \citep{duchene09}.
  
\end{enumerate}

%====================
%Section 3
\section{Close binary formation with a tertiary stellar companion}
\label{sec:close_binary_formation}
%=====================

%----------------------------------------------------
% Figure 2
%
\begin{figure}  
\begin{center}  
	\begin{subfigure}[b]{0.49\textwidth}
		\caption{}
		\includegraphics[width=\textwidth]{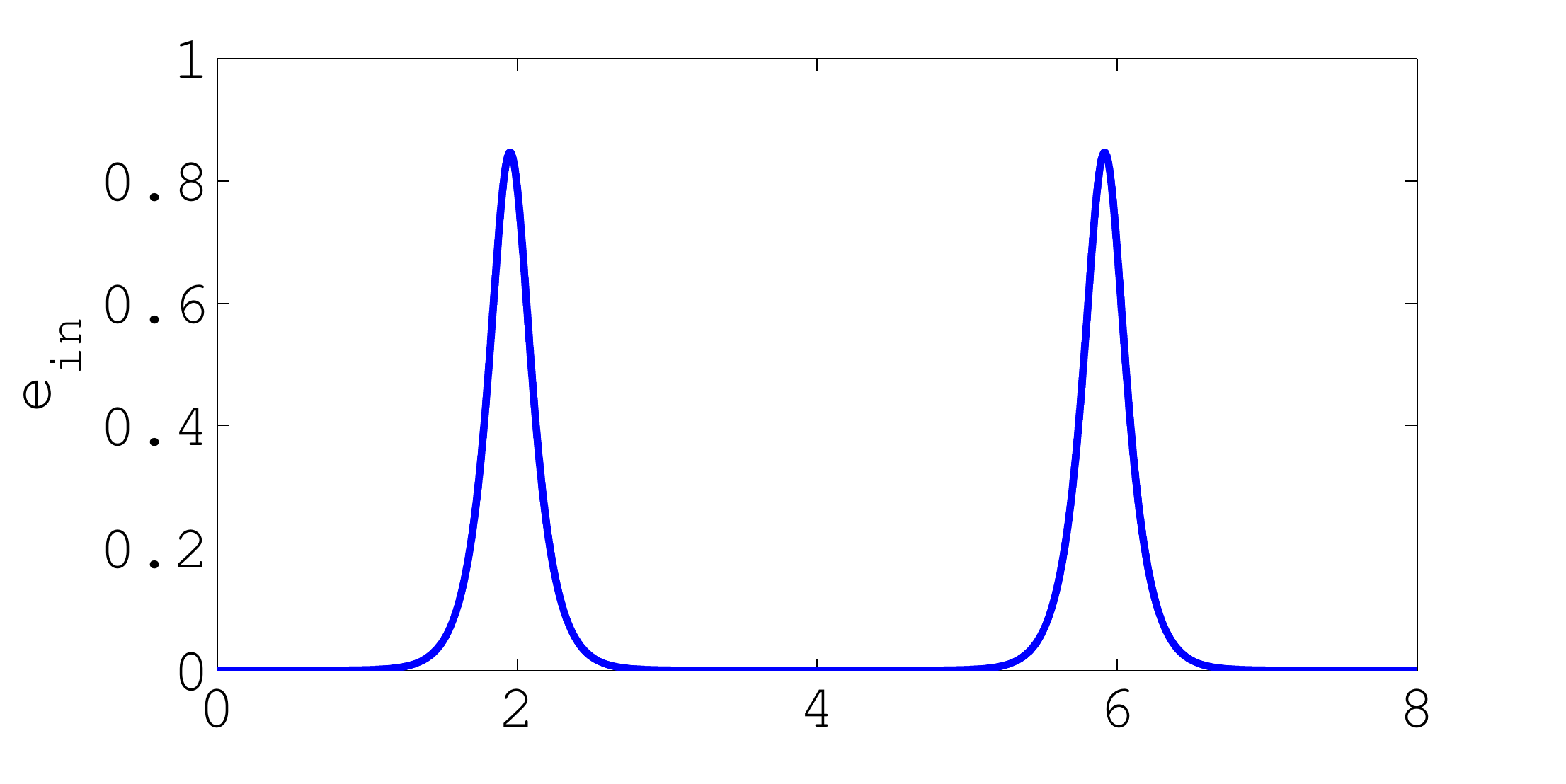}  
		\label{fig:kozai_example_ecc}  
	\end{subfigure}
	\begin{subfigure}[b]{0.49\textwidth}
		\caption{}
		\includegraphics[width=\textwidth]{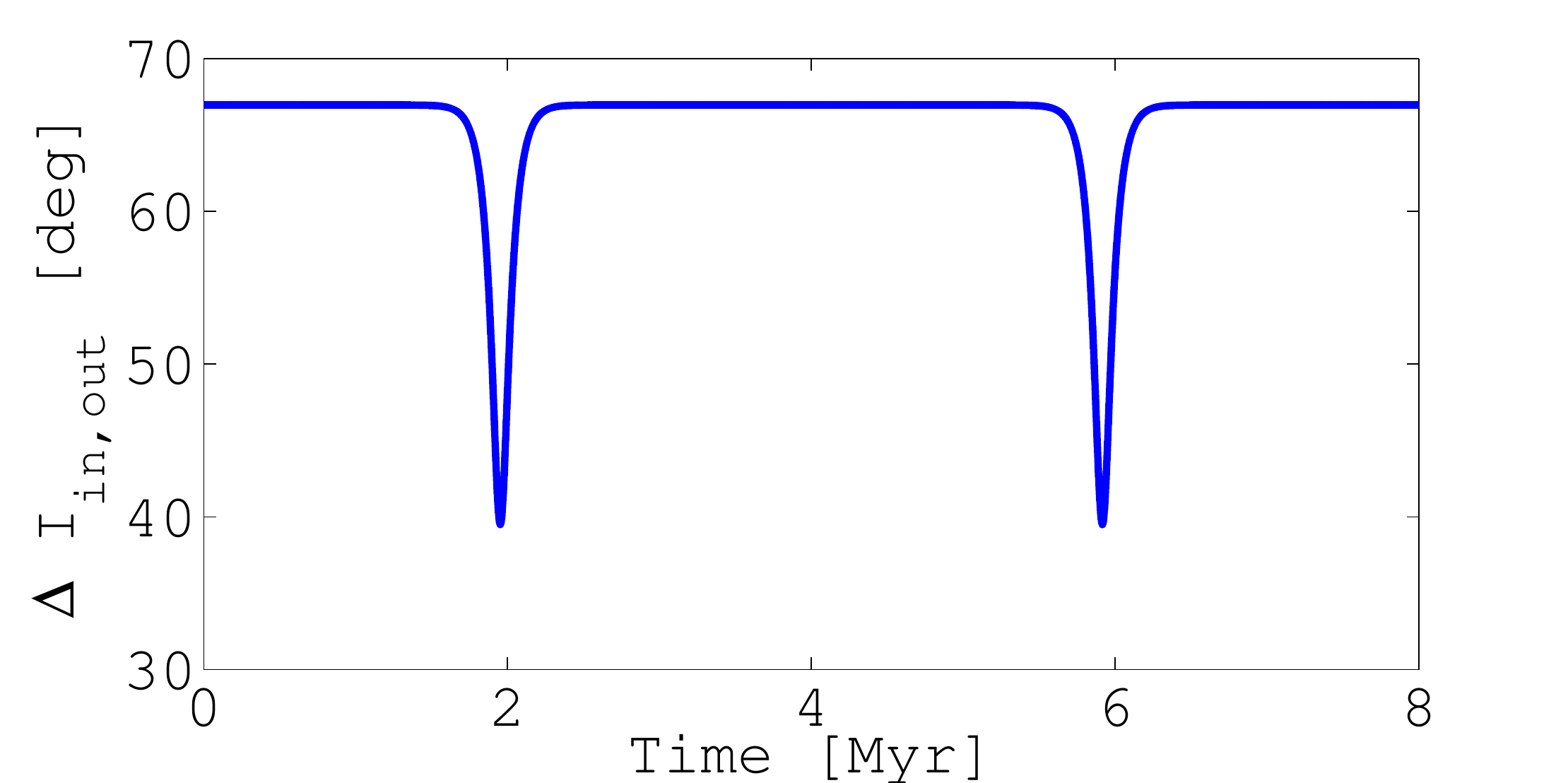}  
		\label{fig:kozai_example_DeltaI}  
	\end{subfigure}	\caption{Example of Kozai cycles of $e_{\rm in}$ (a) and $\Delta I_{\rm in,out}$ (b) for $P_{\rm in} = 100$ d, $P_{\rm out} = 338$ yr, $M_1=M_2=M_3=M_{\odot}$ and $\Delta I_{\rm in,out,init} = 67^{\circ}$. Both the inner and outer binaries are on initially circular orbits.
}\label{fig:kozai_example}  
\end{center}  
\end{figure} 
%--------------------------------------------------------------

A common belief is that short period binaries are not formed in situ, because the fragmenting protostellar cloud contains an excess of angular momentum with respect to the orbital angular momentum of a very close binary (e.g., \citealt{bate12}). These binaries were likely formed much farther apart and subsequently evolved to their present state. An alternate theory for the formation of the closest of binaries ($\lesssim 7$ d) is the combination of tidal interactions and an inclined perturbing tertiary star. This process is canonically known as Kozai cycles with tidal friction (KCTF) and was first proposed by \citet{mazeh79}, and later studied in detail by \citet{eggeleton06}, \citet{fabrycky07} and \citet{naoz14} among others. In this section we summarise this shrinking mechanism.

%---------------------------------------
\subsection{The Kozai modulation}
\label{sec:kozai_binary}
%-------------------------------------

To a first approximation, a hierarchical triple star system can be modelled as an inner binary of two stars and an outer binary composed of the inner binary, 
located at its centre of mass, and the outer tertiary star. Both binaries move on Keplerian orbits which we define using osculating orbital elements for the period, $P$, semi-major axis, $a$, eccentricity, $e$, argument of periapse, $\omega$, and longitude of the ascending node, $\Omega$, where we denote the inner and outer binaries with subscripts ``in" and ``out", respectively. The two orbits are inclined with respect to each other by $\Delta I_{\rm in,out}$. The inner binary stars have masses $M_1$ and $M_2$ and the tertiary mass is $M_3$.

In the quadrupole approximation of the Hamiltonian of the system the tertiary star remains on a static orbit whilst its perturbations induce a nodal and apsidal precession on the inner binary (variations in $\Omega_{\rm in}$ and $\omega_{\rm in}$, respectively). If the two orbital planes are initially misaligned by more than a critical value\footnote{This critical mutual inclination is for an initially circular inner binary. The critical value decreases with increasing initial eccentricity in the inner binary.}, $\Delta I_{\rm in,out,init}>39.2^{\circ}$, there is a variation of $e_{\rm in}$ and $\Delta I_{\rm in,out}$ on the same period as the precession of $\omega_{\rm in}$, and a nodal precession of $\Omega_{\rm in}$ on twice this period (see \citealt{takeda08} for a more in-depth summary). The quadrupole approximation works best when $a_{\rm out}$/$a_{\rm in}$ is very large, and we will be considering systems like this.

An initially circular inner binary reaches an eccentricity of

\begin{equation}
e_{\rm in, max} = \left(1-\frac{5}{3}\cos^2\Delta I_{\rm in,out,init}\right)^{1/2},
\label{eq:kozai_binary_ecc}
\end{equation}
%----------------------------------------
whilst the mutual inclination varies between its initial value and the critical value $39.2^{\circ}$. These variations together constitute the Kozai cycles \citep{lidov62,kozai62}\footnote{ This effect was first published (in Russian) by Lidov but was independently discovered by Kozai later that year. It should arguably be called a Lidov-Kozai cycle but it is more commonly referred to as simply a Kozai cycle, a convention which we follow in this paper.}. The timescale of Kozai cycles is
\begin{equation}
\tau_{\rm Kozai, in} \simeq \frac{2}{3\pi}\frac{P_{\rm out}^2}{P_{\rm in}} \frac{M_1+M_2+M_3}{M_3} \left(1-e_{\rm out}^2\right)^{3/2},
\label{eq:tau_kozai_binary}
\end{equation}
(e.g., \citealt{mazeh79,fabrycky07}). The period of Kozai cycles equals $\tau_{\rm Kozai,in}$ multiplied by a factor of order unity \citep{ford00}. 

A numerical example is given in Fig.~\ref{fig:kozai_example}, demonstrating that $e_{\rm in}$ can be excited to very high values. This simulation, and all other simulations in this paper, was created using the n-body code REBOUND \citep{rein12}\footnote{This easy to use code can be downloaded freely at http://github.com/hannorein/rebound.}. The code uses a 14/15th-order integrator which is not inherently symplectic but uses an adaptive time step to minimise error propagation  generally down to machine precision. Consequently, it preserves the symplectic nature of Hamiltonian systems better than typical symplectic integrators \citep{rein15}. The code is also suitable for high-eccentricity orbits, like those experienced during Kozai cycles. All parameters for the integration were kept at default.

\begin{figure}  
\begin{center}  
\includegraphics[width=0.49\textwidth]{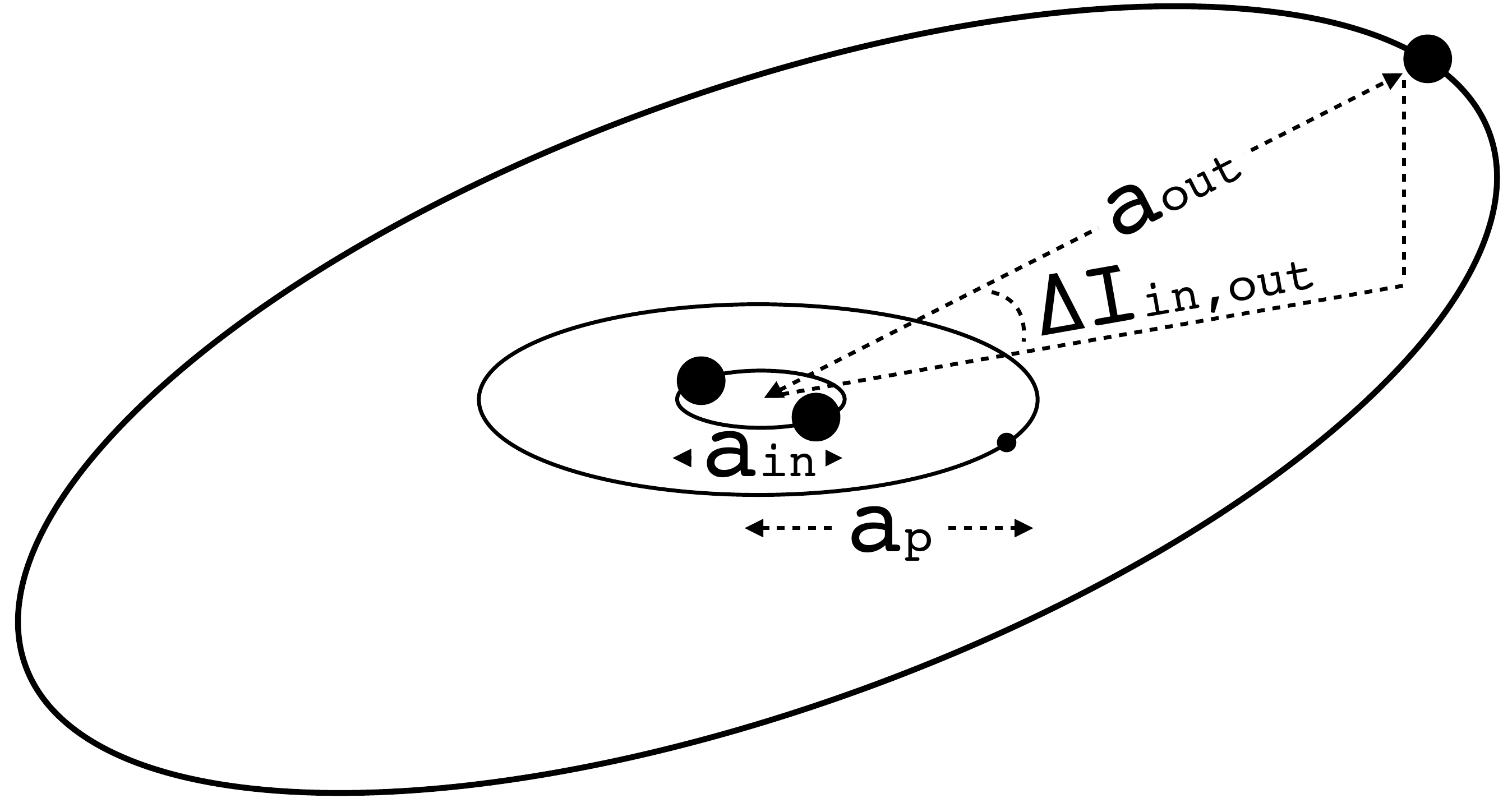} 
\caption{Geometry of a close binary orbited by a planet and a distant misaligned stellar companion. The orbits are drawn with respect to the centre of mass of the inner binary. All orbits are actually circular but misaligned with respect to our line of sight. The planet is coplanar with the inner binary.}
\label{fig:geometry}  
\end{center}  
\end{figure}

%--------------------------------------------------
\subsection{Suppression of Kozai cycles}
\label{sec:suppression}
%---------------------------------------------------

%
Kozai cycles are the result of small perturbations induced by the tertiary star which build up coherently over the apsidal precession period. If there is an additional secular perturbation causing an apsidal precession of the binary on a shorter timescale, then the coherent eccentricity modulation is partially lost. Consequently, the amplitude of the tertiary's perturbations decreases and the Kozai effect is suppressed. 

If the inner binary stars are close enough then apsidal precession due to general relativity and tidal and rotational bulges can suppress the Kozai modulation \citep{wu03,fabrycky07}. Alternatively, a sufficiently close and massive circumbinary planet may also suppress the Kozai modulation by inducing a competing apsidal advance, an effect we consider later in Sect.~\ref{sec:suppression_from_planet} and has been studied by \citet{hamers15a}.

%-----------------------------------------
\subsection{Tidal shrinkage}
%-----------------------------------------

%\begin{equation}
%\tau_{\rm tides} \simeq \tau_{\rm visc}\left[\frac{a_{\rm in}(1-e_{\rm in})}{R_1}\right]^8,
%\label{eq:tides_timescale}
%\end{equation}
%where $\tau_{\rm visc} \sim 1$ yr is the viscous timescale in the star and $R_1$ is the radius of the primary star (REF). The power of 8 in Eq.~\ref{eq:tides_timescale} means that the two stars must be brought very close together for tides to have a significant effect during the lifetime of the binary. 

%
Assuming the Kozai cycles have not been suppressed, the high eccentricity excursions lead to close periapse passages between the two stars of the close binary. During these close encounters there may be a strong tidal interaction, which can dissipate orbital energy that causes the inner binary orbit to both shrink and circularise. The final semi-major axis can be approximated as twice the periapse distance at closest approach,

\begin{equation}
a_{\rm in, final} = 2a_{\rm in,init}(1-e_{\rm in,max}),
\label{eq:final_binary_a}
\end{equation}
although this formula is only valid for when the two stars are close enough for tidal forces to be significant \citep{ford06}.

Observational studies of binary and triple star systems (e.g., \citealt{tokovinin93, tokovinin04, tokovinin06, tokovinin08, tokovinin14a, tokovinin14b}) are consistent with this shrinkage scenario. \citet{tokovinin06} discovered that 96\% of binaries with less than 3 d periods are surrounded by a tertiary stellar companion. This abundance drops to 34\% for binaries with periods longer than 12 d.

%============================
%Section
\section{Planetary dynamics in multi-stellar systems}
\label{sec:planetary_dynamics_multi-stellar}
%============================

There are two possible planetary orbits in binary star systems:

\begin{enumerate}

\item a circumbinary orbit around both stars or
\item a circumprimary orbit around one of the two stars\footnote{Technically a circumprimary orbit only refers to when a planet orbits the bigger of the two stars and an orbit around the smaller star is a circumsecondary orbit, but for simplicity in this paper we will use the term circumprimary to refer to either case.}.

These orbits are also sometimes referred to as p-type and s-type orbits, respectively, but we will not use this nomenclature.

There is an even greater variety of planetary orbits in triple star systems (see \citealt{verrier07} for a classification scheme). In this paper we are only concerned with the scenario illustrated in Fig.~\ref{fig:geometry}, where the planet orbits the inner binary and the third star is on the periphery. 
The planet  may be considered as both on a circumbinary orbit around the inner binary, and on a circumprimary orbit with respect to the outer binary, where the planet orbits a ``star" that is composed of the inner binary. In Sect.~\ref{sec:planets_binaries} we analyse both types of orbits in the context of binary star systems, before combining them in the context of triple stellar systems in Sect.~\ref{sec:planets_triples}.

\end{enumerate}

%====================
%Subsection 
\subsection{Planets in binary star systems}
\label{sec:planets_binaries}
%====================

%====================
%Subsubsection 
\subsubsection{Stability}
\label{sec:planets_binaries_stability}
%====================

There are restrictions on where a planet may orbit stably in a binary star system, primarily as a function of the binary's semi-major axis (e.g., \citealt{dvorak86,holman99}). Planets orbiting on the wrong side of the stability limit are generally ejected from the system by a process of resonance overlap \citep{mudryk06}. Assuming circular {\it circumbinary} orbits \citet{holman99} used an empirical fit to n-body simulations to define the stability limit using a critical semi-major axis,

\begin{align}
a_{\rm crit,CB} &= a_{\rm in}(1.60 + 5.10e_{\rm in}  + 4.12 \mu_{\rm in} - 4.27 e_{\rm in}\mu_{\rm in} \nonumber \\
& \quad - 2.22e_{\rm in}^2 - 5.09 \mu_{\rm in}^2 + 4.61 e_{\rm in}^2\mu_{\rm in}^2),
\label{eq:stability_limit_CB}
\end{align}
where $\mu_{\rm in} = M_2/(M_1+M_2)$ is the mass ratio. We have excluded the fit uncertainties that the authors included in their equation, which are of the order $5\%$. Planets must orbit {\it beyond} this critical semi-major axis, lest they are ejected from the system.

In the alternative case of a circular {\it circumprimary} orbit \citet{holman99} similarly calculated a stability criterion of

\begin{align}
a_{\rm crit,CP} &= a_{\rm out}(0.46  -0.63e_{\rm out}   -0.38 \mu_{\rm out} + 0.59 e_{\rm out}\mu_{\rm out} \nonumber \\
& \quad + 0.15e_{\rm out}^2  -0.20 e_{\rm out}^2\mu_{\rm out}),
\label{eq:stability_limit_CP}
\end{align}
where $\mu_{\rm out} = M_3/(M_1+M_2+M_3)$ is the mass ratio. Equation ~\ref{eq:stability_limit_CP} does not account for eccentric planets. In this configuration the planet must orbit {\it within} this critical semi-major axis in order to maintain stability. Both Eqs.~\ref{eq:stability_limit_CB} and \ref{eq:stability_limit_CP}  were derived for a binary eccentricity between 0 and 0.7--0.8, and a binary mass ratio between 0.1 and 0.9.

The analysis of \citet{holman99} was restricted to coplanar orbits. \citet{doolin11} showed that in the circumbinary case $a_{\rm crit}$ is only a weakly decreasing function with mutual inclination, and that stable circumbinary orbits are possible at all misalignments, including retrograde orbits. 

%====================
%Subsubsection 
\subsubsection{Secular evolution}
%====================

A circumbinary planet experiences a nodal and apsidal precession, of $\omega_{\rm p}$ and $\Omega_{\rm p}$, respectively, both of which occur at approximately the same rate but in opposite directions (e.g., \citealt{lee06}). The timescale of this precession is

\begin{equation}
\tau_{\rm prec, p} \simeq \frac{4}{3}\left(\frac{P_{\rm p}^7}{P_{\rm in}^4}\right)^{1/3}\frac{(M_1+M_2)^2}{M_1M_2}\frac{\left(1-e_{\rm p}^2\right)^2}{\cos\Delta I_{\rm p, in}},
%(4/3)*((P_p*days2sec)^7/(P_b*days2sec)^4)^(1/3)*m1*m2/(m1+m2)^2*(1-e_p^2)^(3/2)/cos(deltaI);
\label{eq:tau_prec}
\end{equation}
which is adapted from \citet{farago10} to be in terms of orbital periods. For circular binaries the mutual inclination $\Delta I$ remains constant, but for eccentric binaries it may vary slightly \citep{farago10,doolin11}. 

A circumprimary orbit will also experience a precession $\omega_{\rm p}$ and $\Omega_{\rm p}$. If the orbit is misaligned by at least  $39.2^{\circ}$ with respect to the outer binary it will undergo Kozai cycles, as was seen for stellar binaries in Sect.~\ref{sec:kozai_binary}, with a timescale
\begin{equation}
\tau_{\rm Kozai, p} \simeq \frac{2}{3\pi}\frac{P_{\rm out}^2}{P_{\rm p}} \frac{M_1+M_{\rm p}+M_2}{M_2} \left(1-e_{\rm out}^2\right)^{3/2}.
\label{eq:tau_kozai_planet}
\end{equation}
Note that Kozai cycles only apply to circumprimary orbits, not to circumbinary ones.

%====================
%Subsection 
\subsection{Planets in triple star systems}
\label{sec:planets_triples}
%====================

%====================
%Subsubsection 
\subsubsection{Stability}
%====================

\begin{figure}  
\begin{center}  
\includegraphics[width=0.49\textwidth]{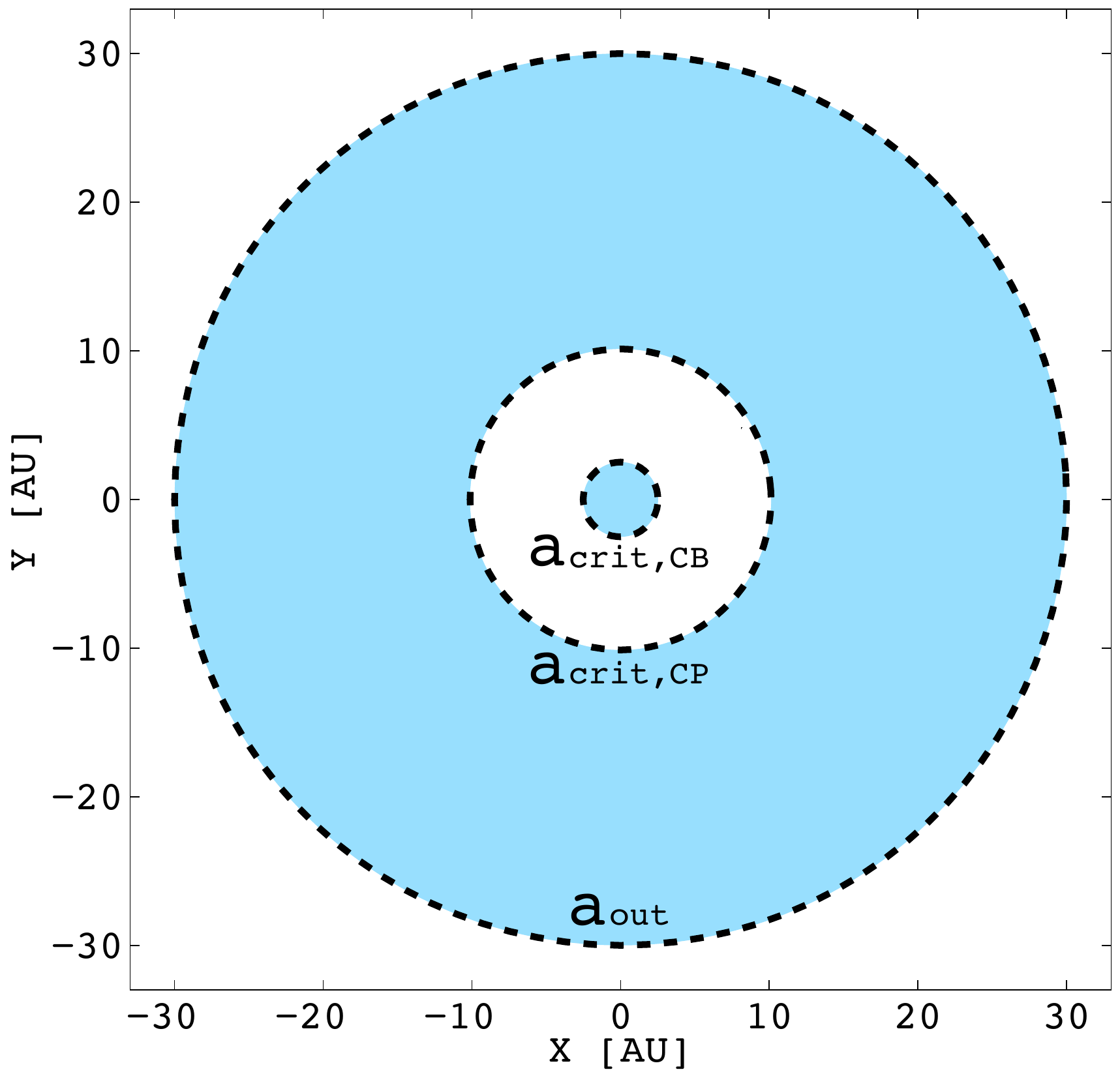} 
\caption{Stability limits in a triple star system from Eqs.~\ref{eq:stability_limit_CB} and ~\ref{eq:stability_limit_CP}, for $M_1=M_2=M_3=M_{\odot}$, $a_{\rm in} = $ 0.5 AU, $a_{\rm out} = 30$ AU, $e_{\rm in} = e_{\rm out} = 0$ and all orbits are coplanar. The two blue shaded regions are unstable for planetary orbits and the white region in between is stable. Note that this does not take into account eccentric planets.}
\label{fig:double_stability_limit}  
\end{center}  
\end{figure} 

Because a planetary orbit in a triple star system has both circumbinary and circumprimary characteristics, it must obey both stability constraints $a_{\rm crit,CB}$ (Eq.~\ref{eq:stability_limit_CB}) and $a_{\rm crit,CP}$ (Eq.~\ref{eq:stability_limit_CP}). If $a_{\rm out}/a_{\rm in}$ is sufficiently large then $a_{\rm crit,CP} > a_{\rm crit,CB}$, and hence there is a region of stable planetary orbits. In Fig.~\ref{fig:double_stability_limit} we illustrate these combined stability limits, using a simple example of $a_{\rm in} = 0.5$ AU, $a_{\rm out} = 30$ AU, equal mass stars and coplanar and circular orbits. 

\citet{verrier07} analysed this scenario with coplanar and circular orbits and concluded that this joint application of Eqs.~\ref{eq:stability_limit_CB} and \ref{eq:stability_limit_CP} is a valid description of planetary stability in triple star systems. This implies that the perturbations from the inner and outer binaries act somewhat independently. This assumption was seen to become less valid when the inner and outer binary orbits were made eccentric, resulting in a smaller stability region than what is defined by Eq.~\ref{eq:stability_limit_CB} and Eq.~\ref{eq:stability_limit_CP}. This additional instability was attributed to combined perturbations from the ensemble of three stars. \citet{verrier07} did not test planet eccentricity, which would no doubt make the stability constraints even more restrictive.

%====================
%Subsubsection 
\subsubsection{Secular evolution}
\label{sec:planets_triples_secular}
%====================

A planet orbiting within the stability region of a misaligned triple system is perturbed by two competing secular effects: precession due to the inner binary and Kozai cycles due to the outer binary. Kozai cycles in the planet may be suppressed by the precession induced by the inner binary, like what was discussed in the case of triple star systems in Sect.~\ref{sec:suppression}. An approximate criterion for suppression of Kozai cycles is

\begin{equation}
\tau_{\rm prec,p} < \tau_{\rm Kozai,p}.
\label{eq:kozai_supression_criterion_1}
\end{equation}
%which by combining Eqs.~\ref{eq:tau_prec} and \ref{eq:tau_kozai_planet} becomes
%\begin{equation}
%P_{\rm p} < \left[\frac{P_{\rm in}^{4/3}P_{\rm out}^2}{2\pi}\frac{\left(M_1+M_2\right)^2\left(M_1+M_2+M_3\right)}{M_1M_2M_3}\frac{\left(1-e_{\rm out}\right)^{3/2}}{\left(1-e_{\rm p}^2\right)^2}\cos\Delta I_{\rm p,in}\right]^{3/10}.
%\label{eq:kozai_supression_criterion_2}
%\end{equation}
%By taking equal mass stars, circular orbits and $\Delta I_{\rm p,in} = 0$ the criterion becomes
%
%\begin{equation}
%P_{\rm p} < \left[\frac{6}{\pi}P_{\rm in}^{4/3}P_{\rm out}^2\right]^{3/10}.
%\label{eq:kozai_supression_criterion_3}
%\end{equation}
In reality there is likely to be a transition between the two competing secular effects when their timescales are similar. \citet{verrier09} showed that the inner binary can ``protect" the planet through this suppression of Kozai cycles. On the other hand, planets left to undergo high-amplitude Kozai cycles were seen by \citet{verrier09} to become unstable rapidly. Some alternative means of suppressing Kozai cycles, such as general relativistic precession and tidal and rotational bulges in the stars, are not applicable here. This is because these effects are only non-negligible for planets very close to the stars, and such orbits are prohibitively unstable for circumbinary planets. The Kozai effect may also be suppressed by perturbations from an additional planet, but we do not consider multi-planetary systems in this analysis (see Sect.~\ref{sec:multi_planet_systems} for a short discussion on the effects of additional planets).

We have not followed the analytic derivation to the octopole order, which can lead to secular resonance between different precession terms resulting in $\sim 0.1$ eccentricity fluctuations over semi-major axis ranges of $\sim 10\%$ \citep{naoz13}.  This effect may lead to additional unstable regions in the CBP parameter space.

As the binary orbit shrinks under KCTF the planet precession timescale increases as $\tau_{\rm prec,p} \propto P_{\rm in}^{-4/3}$, whilst $\tau_{\rm Kozai,p}$ remains constant. This means that the relative influence of Kozai cycles on the planet will increase and may start to dominate.

\begin{figure}  
\begin{center}  
\includegraphics[width=0.49\textwidth]{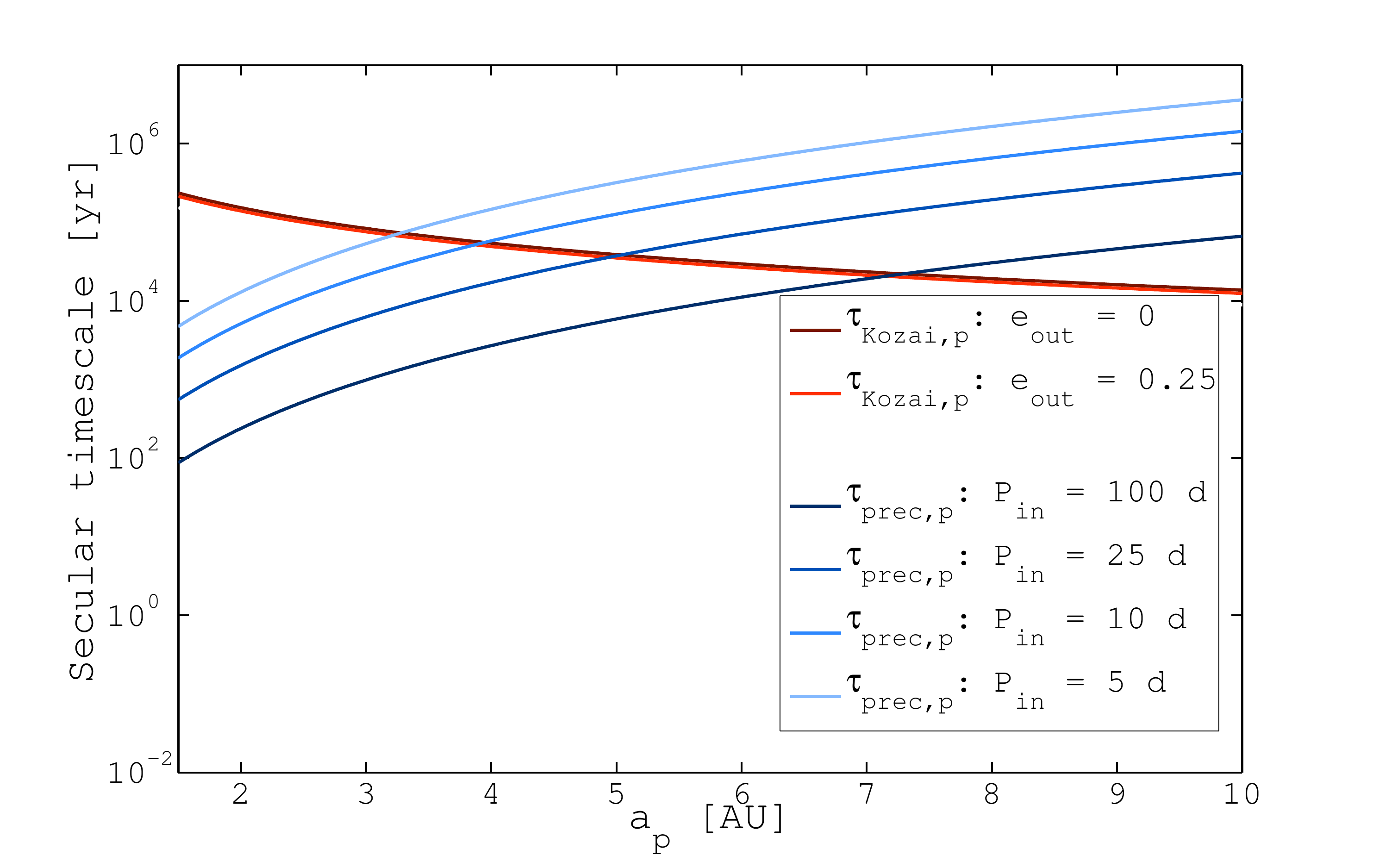} 
\caption{Competing secular timescales on the planet: circumbinary precession (Eq.~\ref{eq:tau_prec}), induced by the inner binary on the planet, and Kozai cycles (Eq.~\ref{eq:tau_kozai_planet}), induced by the third star at 100 AU. The circumbinary precession timescale is shown in different shades of blue for different values of $P_{\rm in}$, starting from 100 d (dark blue) to 5 d (light blue). The Kozai timescale is shown for $e_{\rm out}$ = 0 (maroon) and 0.25 (red) and is independent of $P_{\rm in}$. This plot covers the simulated range of $a_{\rm p}$ between 1.5 and 10 AU.}
\label{fig:competing_timescales}  
\end{center}  
\end{figure}

%====================
%Subsubsection 
\subsubsection{KCTF suppression due to planetary mass}
\label{sec:suppression_from_planet}
%====================
 
Just as Kozai cycles of the planet may be suppressed by the 
apsidal motion induced by the host binary, 
the inner binary might fail to excite to high eccentricity as a result of apsidal precession due to the mass of the planet.  
In effect, if the planet's tidal force exceeds that of the tertiary star, the host binary precesses too quickly for the tertiary to secularly excite its eccentricity.   This is possible even for small planetary masses, because the planet is much closer to the binary than the third star is, and tidal effects scale as separation cubed. 

For Kozai cycles analysed to quadrupole order, we follow \cite{fabrycky07} Sect. 3.2 to compute the maximum eccentricity of the binary in the presence of additional precession.  For a triple system, there are two constants of motion: 
\begin{eqnarray}
F' &=& -2 -3 e^2 + (3+12e^2-15e \cos^2\omega) \sin^2 \Delta I \ , \\
H' &=& (1-e^2)^{1/2} \cos \Delta I \ ,
\end{eqnarray}
where the former is a non-dimensionalised form of the Hamiltonian term defining the interaction between the two orbits, and the latter is a non-dimensionalised form of the z-component of angular momentum\footnote{This corrects a misprint in equation~17 of \cite{fabrycky07}. }, which is the component of the angular momentum of the binary along the total angular momentum of the system. In this context, $e$ and $\omega$ are for inner binary and $\Delta I$ is the mutual inclination of the two orbits.

In the case of a quadruple system, we may consider each level of the hierarchy as an orbit of a body around the centre of mass of the interior bodies.  Averaging over the orbital trajectories gives the following averaged Hamiltonian (e.g., \citealt{ford00}): 
\begin{eqnarray}
\langle \mathcal F \rangle &=& \mathcal F_{\rm in} + \mathcal F_{\rm p} + \mathcal F_{\rm out} +\langle \mathcal F_{\rm in,p} \rangle +\langle \mathcal F_{\rm p,out} \rangle +\langle \mathcal F_{\rm in,out} \rangle,  \\
\langle \mathcal F_{\rm in} \rangle &=& -\frac{G M_1 M_2}{2 a_{\rm in}} \label{eq:fin}, \\
\langle \mathcal F_{\rm p} \rangle &=&-\frac{G (M_1+M_2)M_{\rm p}}{2 a_{\rm p}}, \\
\langle \mathcal F_{\rm out} \rangle &=& -\frac{G (M_1+M_2 +M_{\rm p})M_3}{2 a_{\rm out}} \label{eq:fout},\\
\langle \mathcal F_{\rm in,p} \rangle &=& -\frac{G M_1 M_2 M_{\rm p}}{M_1+M_2} \frac{a_{\rm in}^2}{8 a_{\rm out}^3 (1-e_{\rm out}^2)^{3/2}} \\
&\times& \Big{(} 2 + 3e_{\rm in}^2 - (3+12e_{\rm in}^2-15e_{\rm in}^2 \cos^2 \omega_{\rm in,p} ) \sin^2 \Delta I_{\rm in,p} \Big{)} \nonumber \label{eq:Finp},\\
\langle \mathcal F_{\rm p,out} \rangle &=& -\frac{G (M_1+M_2) M_{\rm p} M_3}{(M_1+M_2+M_{\rm p}} \frac{a_{\rm p}^2}{8 a_{\rm p}^3 (1-e_{\rm p}^2)^{3/2}} \\
&\times& \Big{(} 2 + 3e_{\rm p}^2 - (3+12e_{\rm p}^2-15e_{\rm p}^2 \cos^2 \omega_{\rm p} ) \sin^2 \Delta I_{\rm p,out} \Big{)} \nonumber\label{eq:Fpout}, \\
\langle \mathcal F_{\rm in,out} \rangle &=& -\frac{G M_1 M_2 M_3}{M_1+M_2} \frac{a_{\rm in}^2}{8 a_{\rm out}^3 (1-e_{\rm out}^2)^{3/2}} \\
&\times& \Big{(} 2 + 3e_{\rm in}^2 - (3+12e_{\rm in}^2-15e_{\rm in}^2 \cos^2 \omega_{\rm in} ) \sin^2 \Delta I_{\rm in,out} \Big{)}\nonumber\label{eq:Finout}.
\end{eqnarray}

Suppose the binary begins on a circular orbit, with a coplanar exterior planet ($\Delta I_{\rm p,in} = 0$). The planet's orbit continues on a nearly coplanar orbit, for the same reason that its Kozai cycles are suppressed, as just described in Sect.~\ref{sec:planets_triples_secular}.  We no longer have conservation of $F'$, but rather conservation of $\langle \mathcal F \rangle$.  In the secular problem, without tidal dissipation, $a_{\rm in}$, $a_{\rm p}$, and $a_{\rm out}$ are all conserved, so the Keplerian Hamiltonian terms, Eqs.~\ref{eq:fin} --\ref{eq:fout}, are individually conserved.  The other sub-components can trade energy.

\begin{figure*}  
\captionsetup[subfigure]{labelformat=empty}
\begin{center}  
	\begin{subfigure}[b]{0.49\textwidth}
		\subcaption{{\large$a_{\rm p} = 1.5 $ AU}}
		\includegraphics[width=\textwidth]{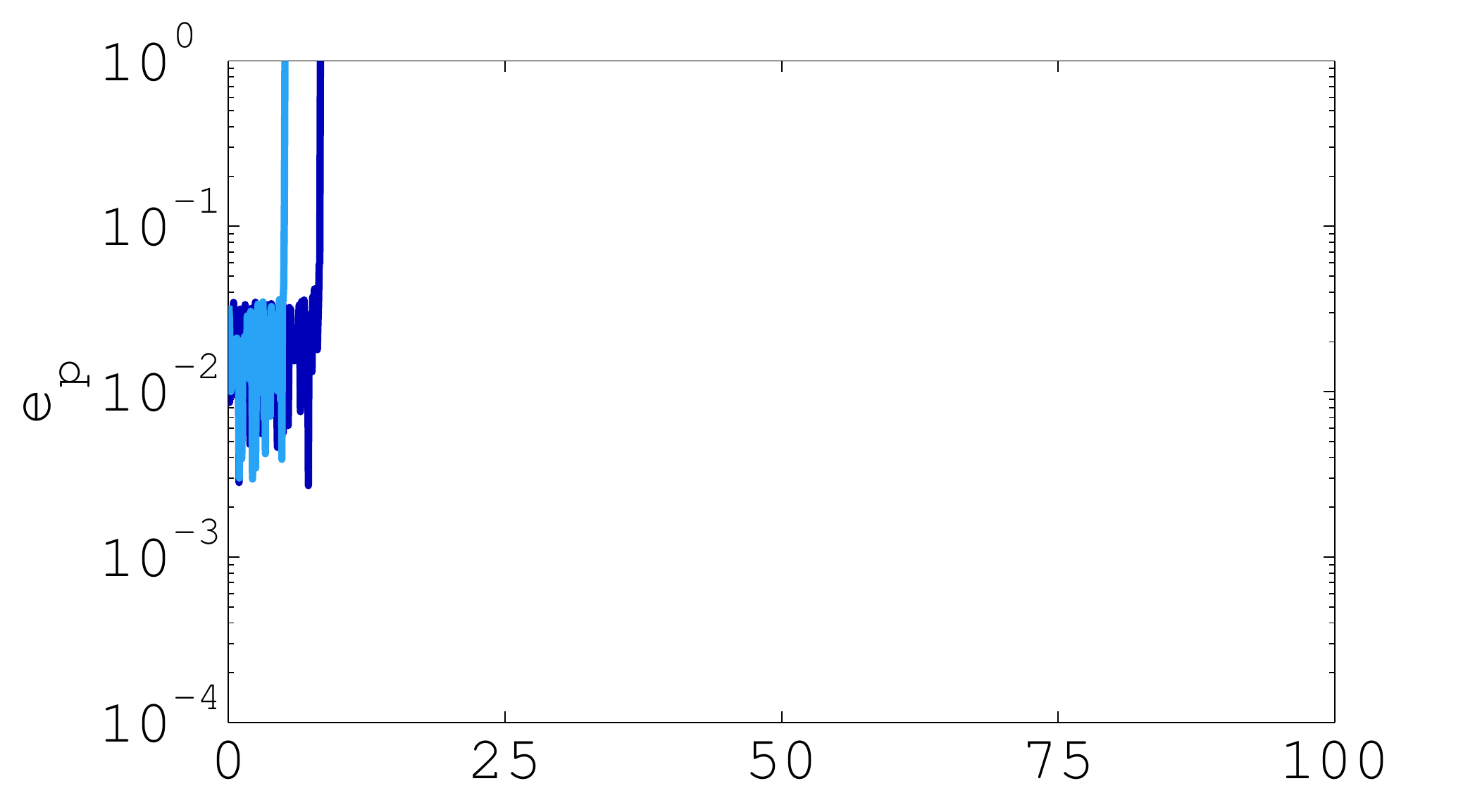}  
	\end{subfigure}
	\vspace{3.5mm}
	\begin{subfigure}[b]{0.49\textwidth}
		\subcaption{{\large$a_{\rm p} = 2 $ AU}}
		\includegraphics[width=\textwidth]{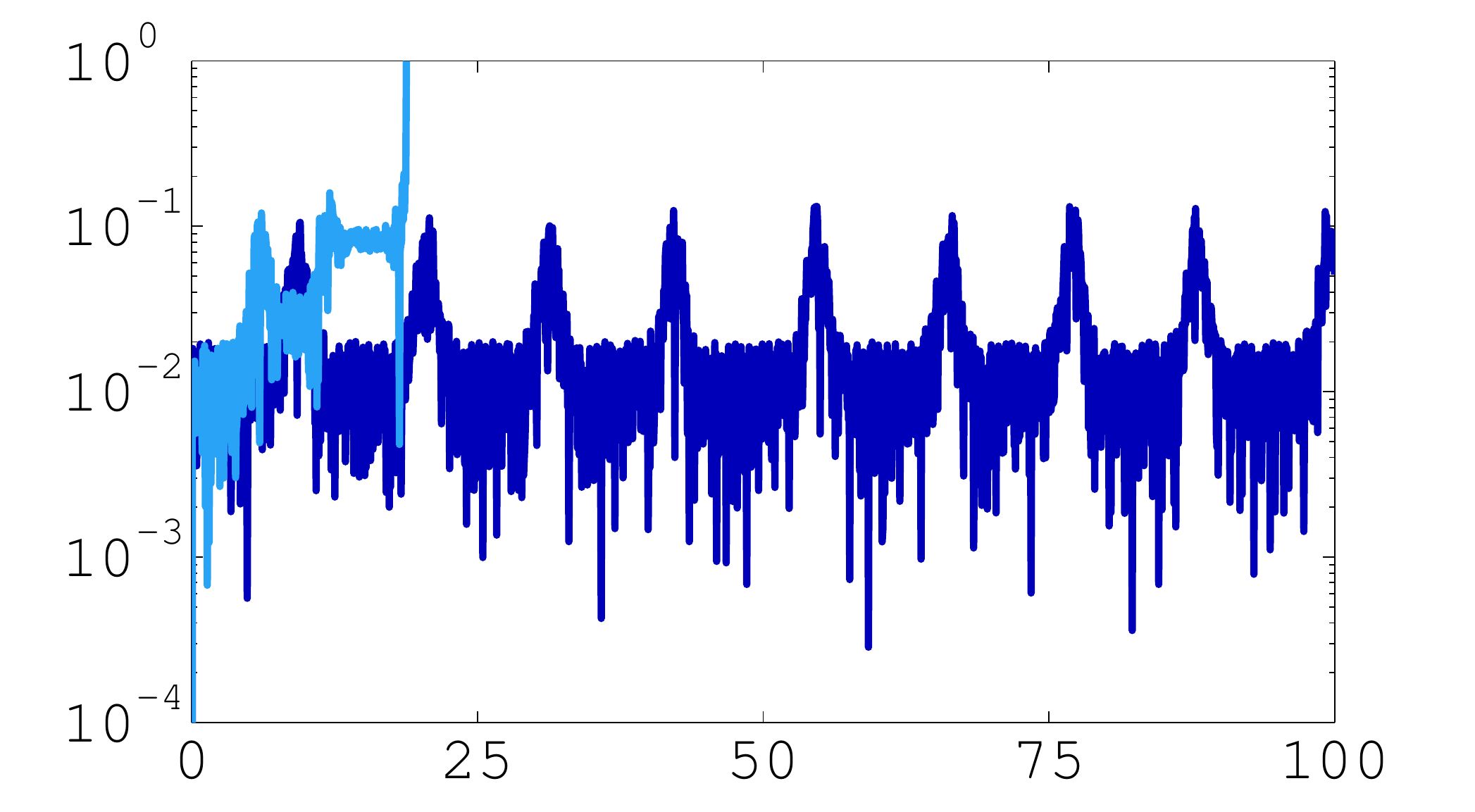}  
	\end{subfigure}	
	\begin{subfigure}[b]{0.49\textwidth}
		\subcaption{{\large$a_{\rm p} = 3.5 $ AU}}
		\includegraphics[width=\textwidth]{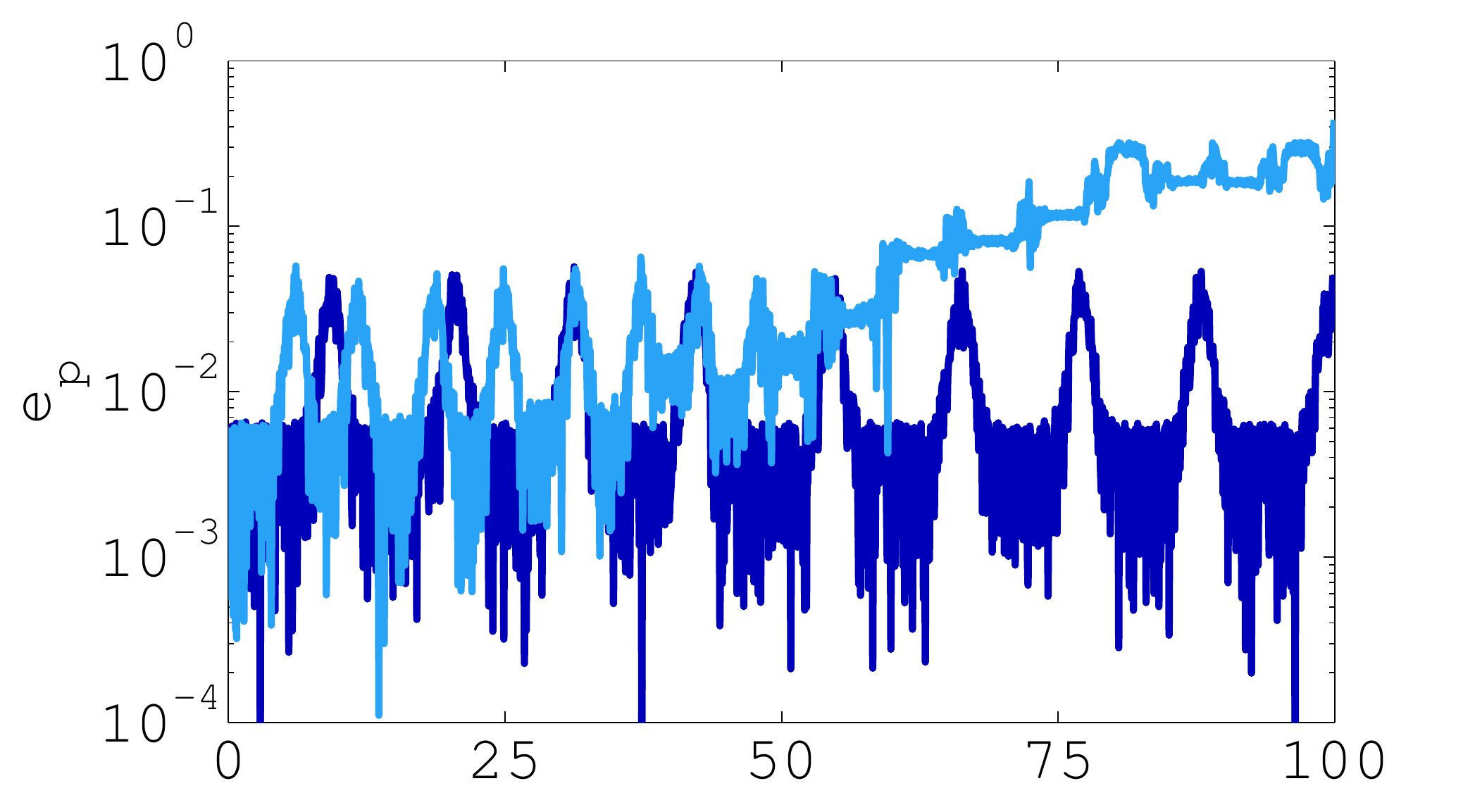}  
	\end{subfigure}
	\vspace{3.5mm}
	\begin{subfigure}[b]{0.49\textwidth}
		\subcaption{{\large$a_{\rm p} = 4 $ AU}}
		\includegraphics[width=\textwidth]{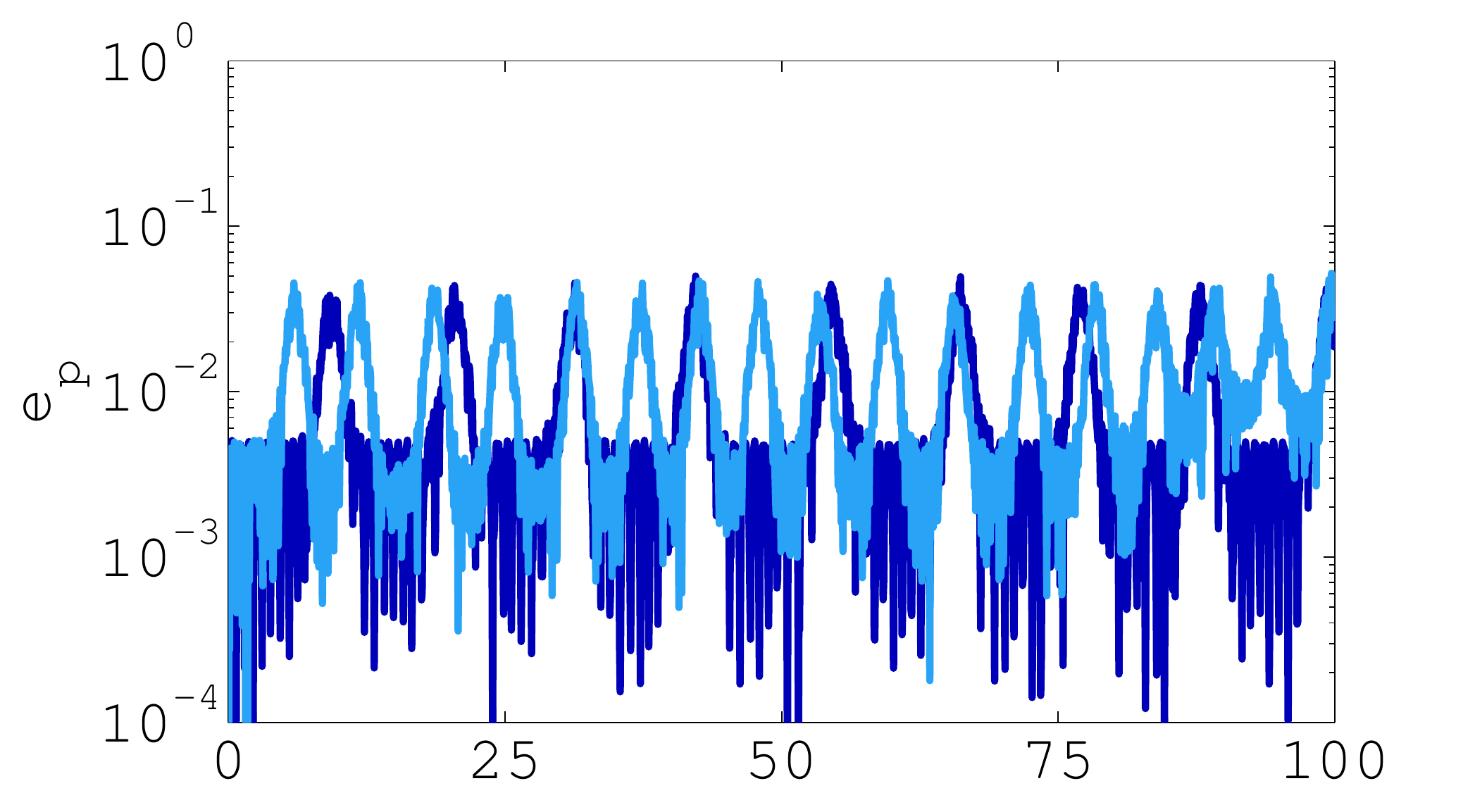}  
	\end{subfigure}
	\begin{subfigure}[b]{0.49\textwidth}
		\subcaption{{\large$a_{\rm p} = 7 $ AU}}
		\includegraphics[width=\textwidth]{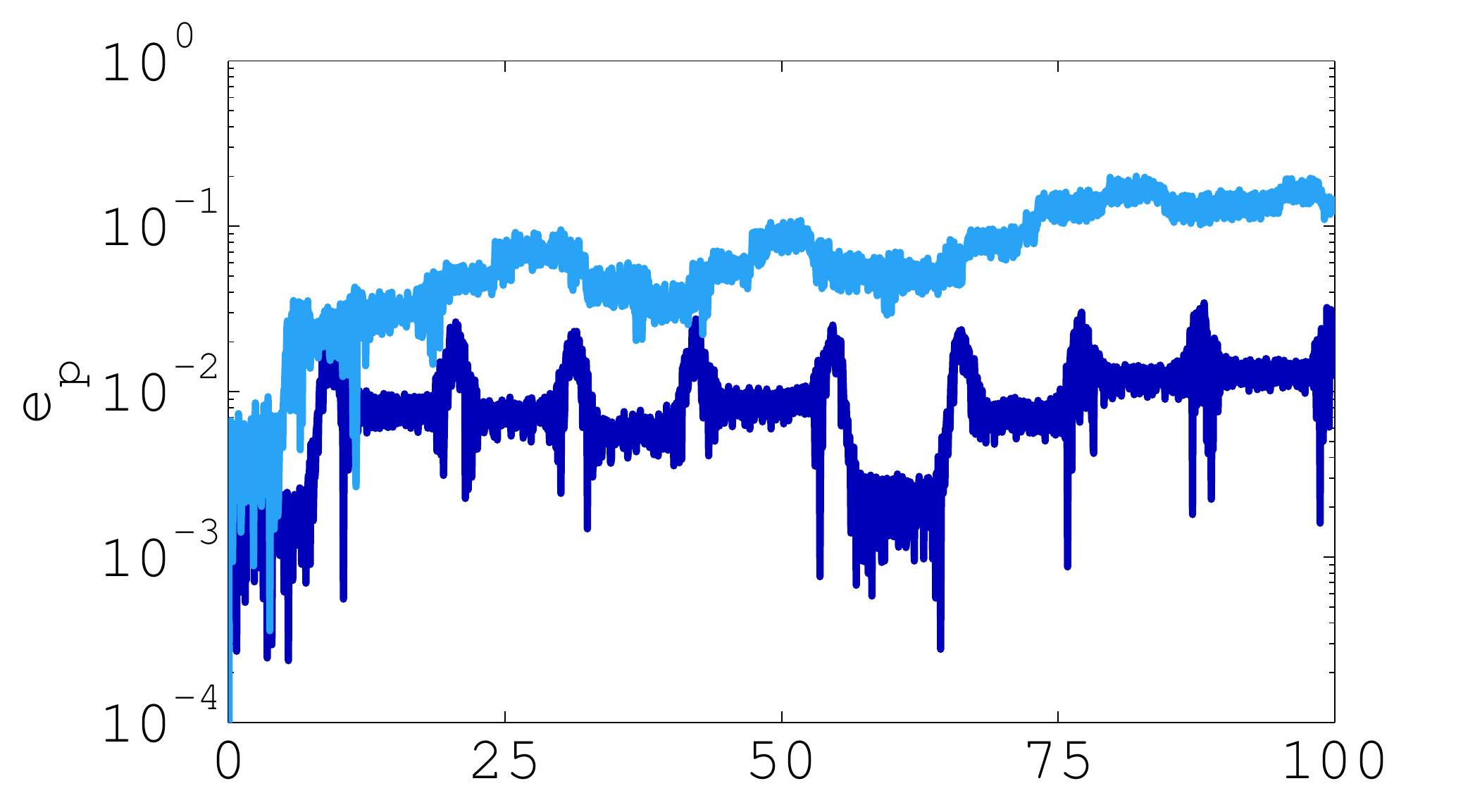}  
	\end{subfigure}
	\vspace{3.5mm}
	\begin{subfigure}[b]{0.49\textwidth}
		\subcaption{{\large$a_{\rm p} = 8 $ AU}}
		\includegraphics[width=\textwidth]{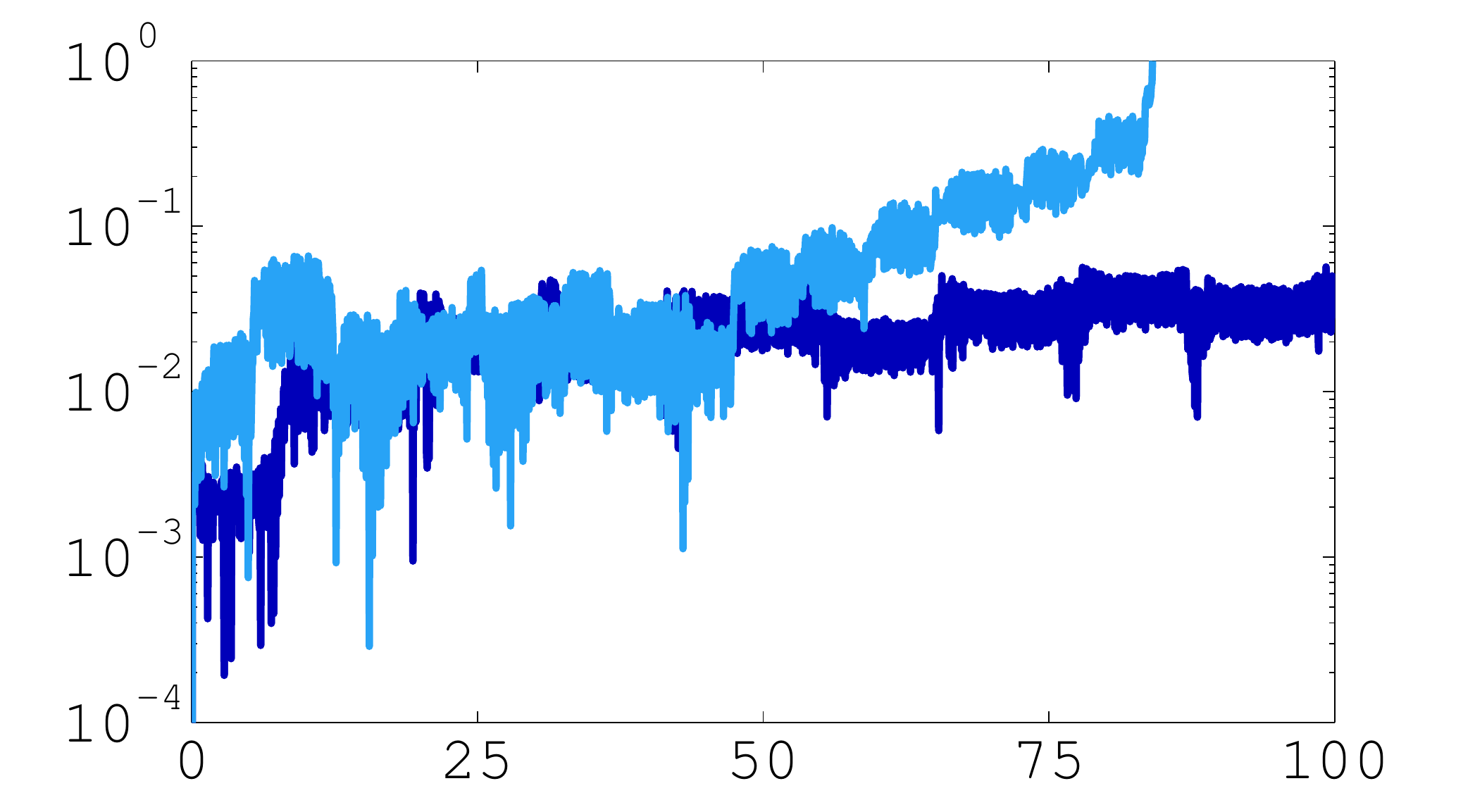}  
	\end{subfigure}
	\begin{subfigure}[b]{0.49\textwidth}
		\subcaption{{\large$a_{\rm p} = 9 $ AU}}
		\includegraphics[width=\textwidth]{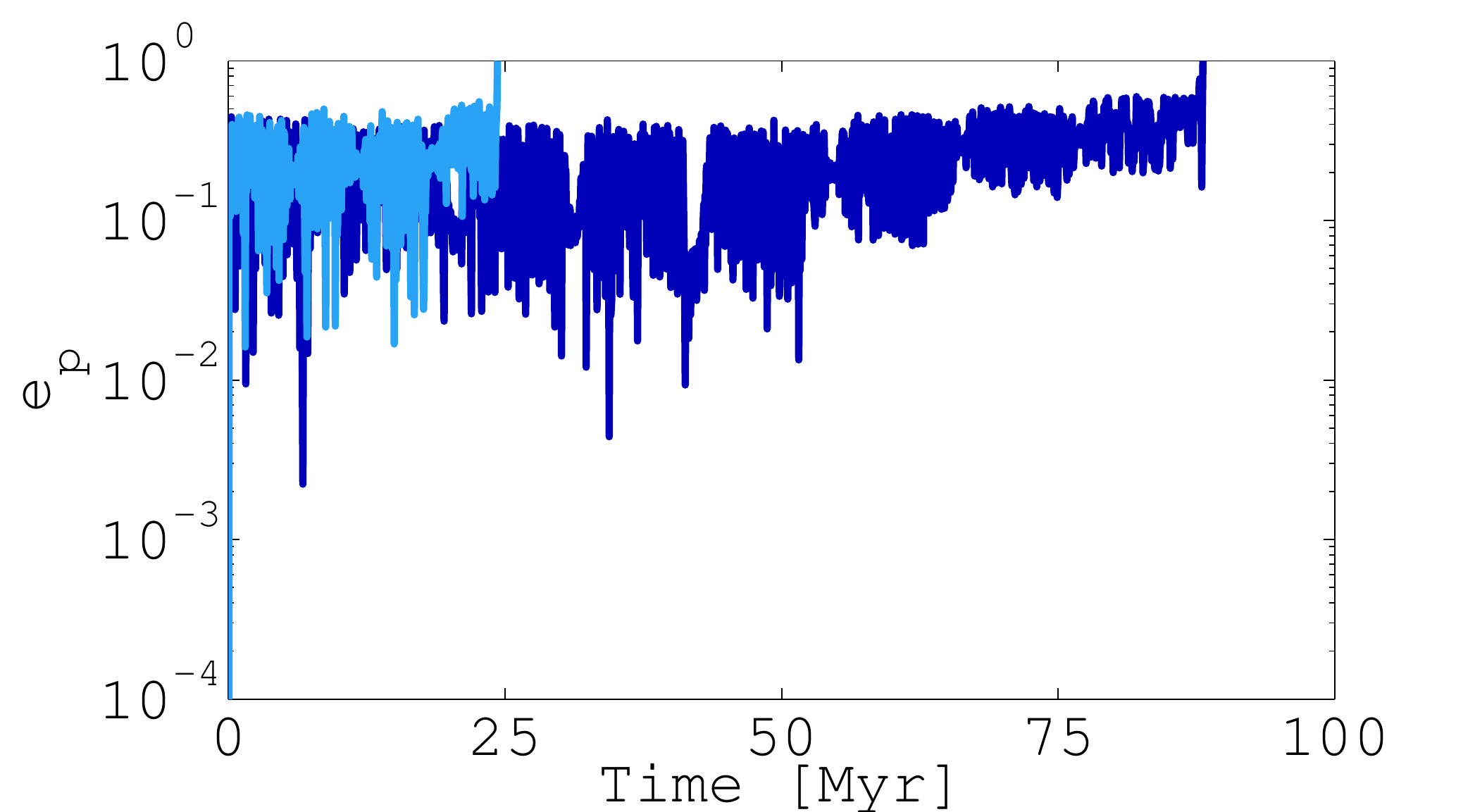}
	\end{subfigure}
	\begin{subfigure}[b]{0.49\textwidth}
		\subcaption{{\large Inner binary}}
		\includegraphics[width=\textwidth]{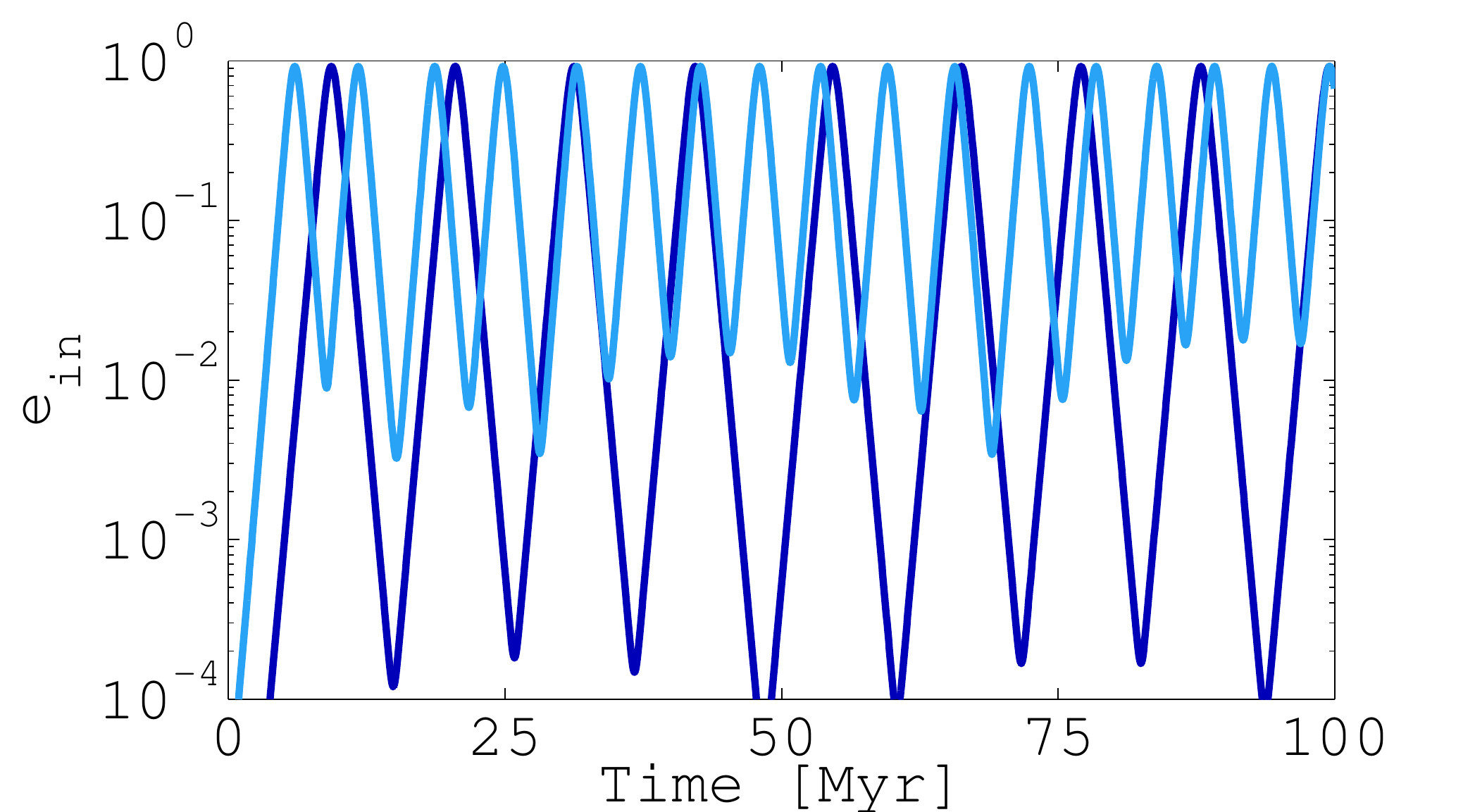}  
	\end{subfigure}
	\caption{ First seven figures: evolution over 100 Myr of the planet eccentricity in a primordial triple star system, where $a_{\rm in} = 0.4827$ AU ($P_{\rm in} = 100$ d), $a_{\rm out} = 100$ AU ($P_{\rm out} = 2.52  \times 10^5$ d) and $e_{\rm out} =$ 0 (dark blue) and 0.25 (light blue) and the inner binary starts on a circular orbit. The planet is massless and starting on an orbit that is circular and coplanar with the inner binary. Each subplot shows a different value of $a_{\rm p}$ (see Table.~\ref{tab:PrimordialParams} for all the simulation parameters). Bottom right figure: evolution of the inner binary eccentricity for $e_{\rm out} =$ 0 (dark blue) and 0.25 (light blue).
}\label{fig:example_eccentricity}  
\end{center}  
\end{figure*} 

\begin{figure*}  
\captionsetup[subfigure]{labelformat=empty}
\begin{center}  
	\begin{subfigure}[b]{0.49\textwidth}
		\subcaption{{\large$a_{\rm p} = 1.5 $ AU}}
		\includegraphics[width=\textwidth]{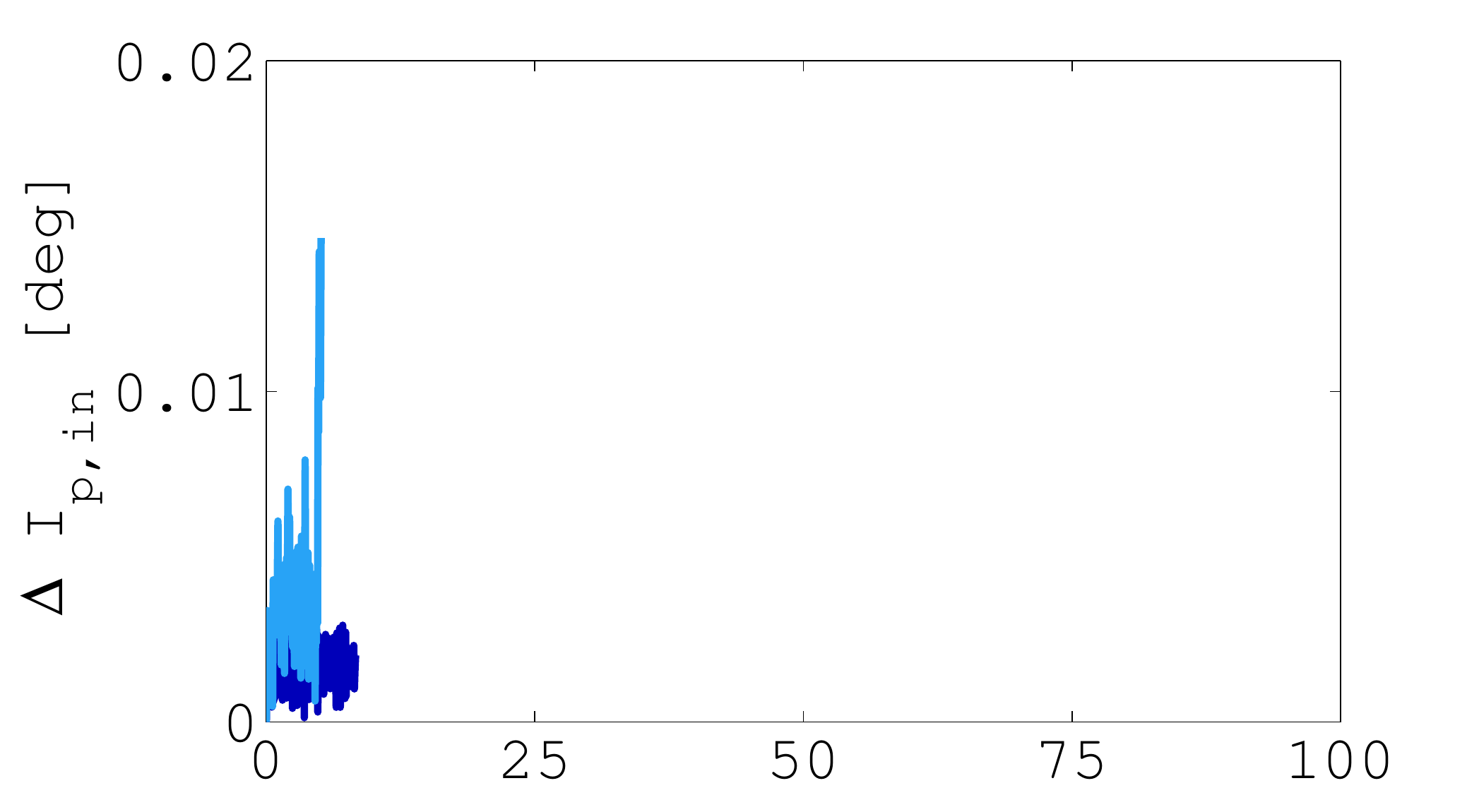}  
	\end{subfigure}
	\vspace{3.5mm}
	\begin{subfigure}[b]{0.49\textwidth}
		\subcaption{{\large$a_{\rm p} = 2 $ AU}}
		\includegraphics[width=\textwidth]{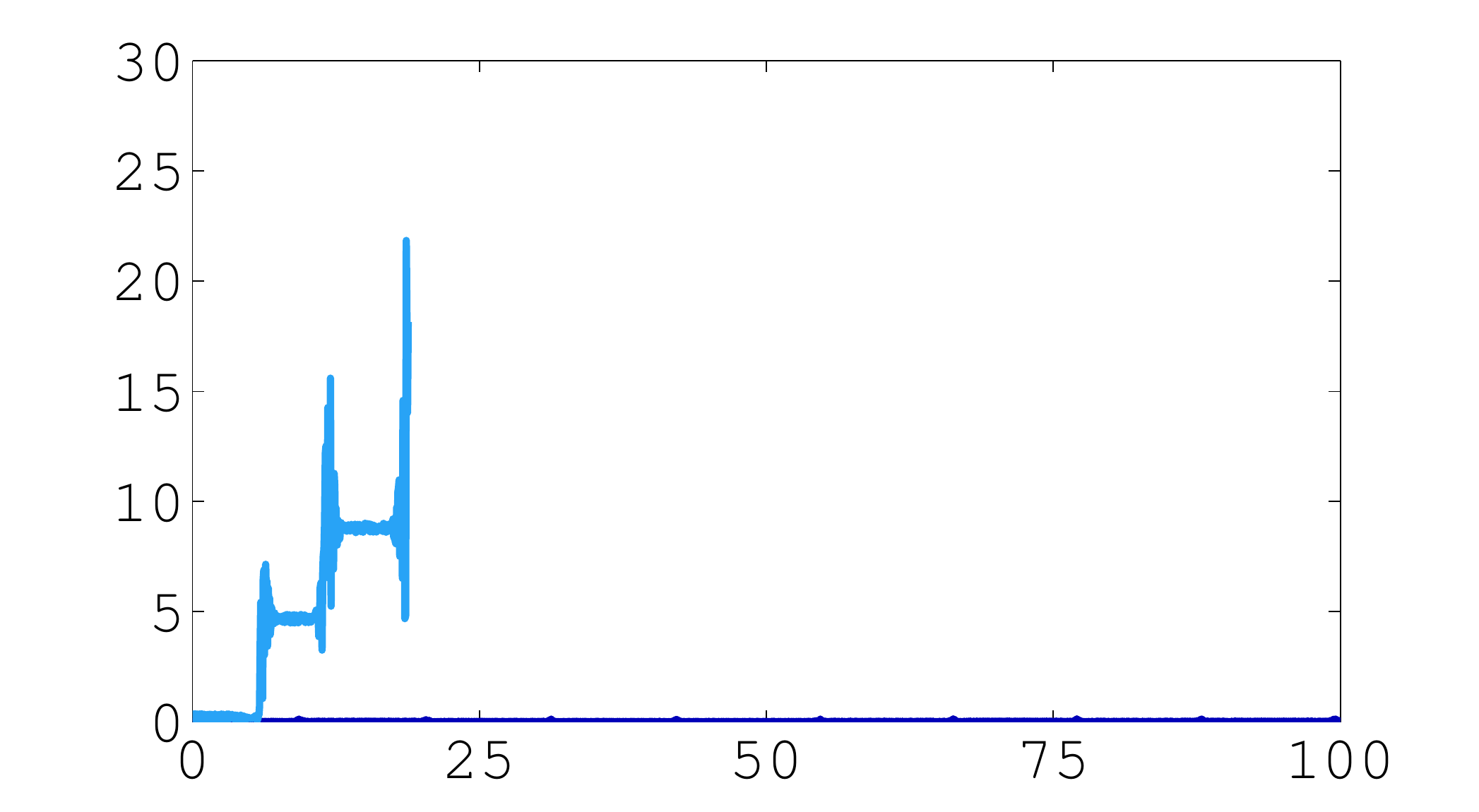}  
	\end{subfigure}	
	\begin{subfigure}[b]{0.49\textwidth}
		\subcaption{{\large$a_{\rm p} = 3.5 $ AU}}
		\includegraphics[width=\textwidth]{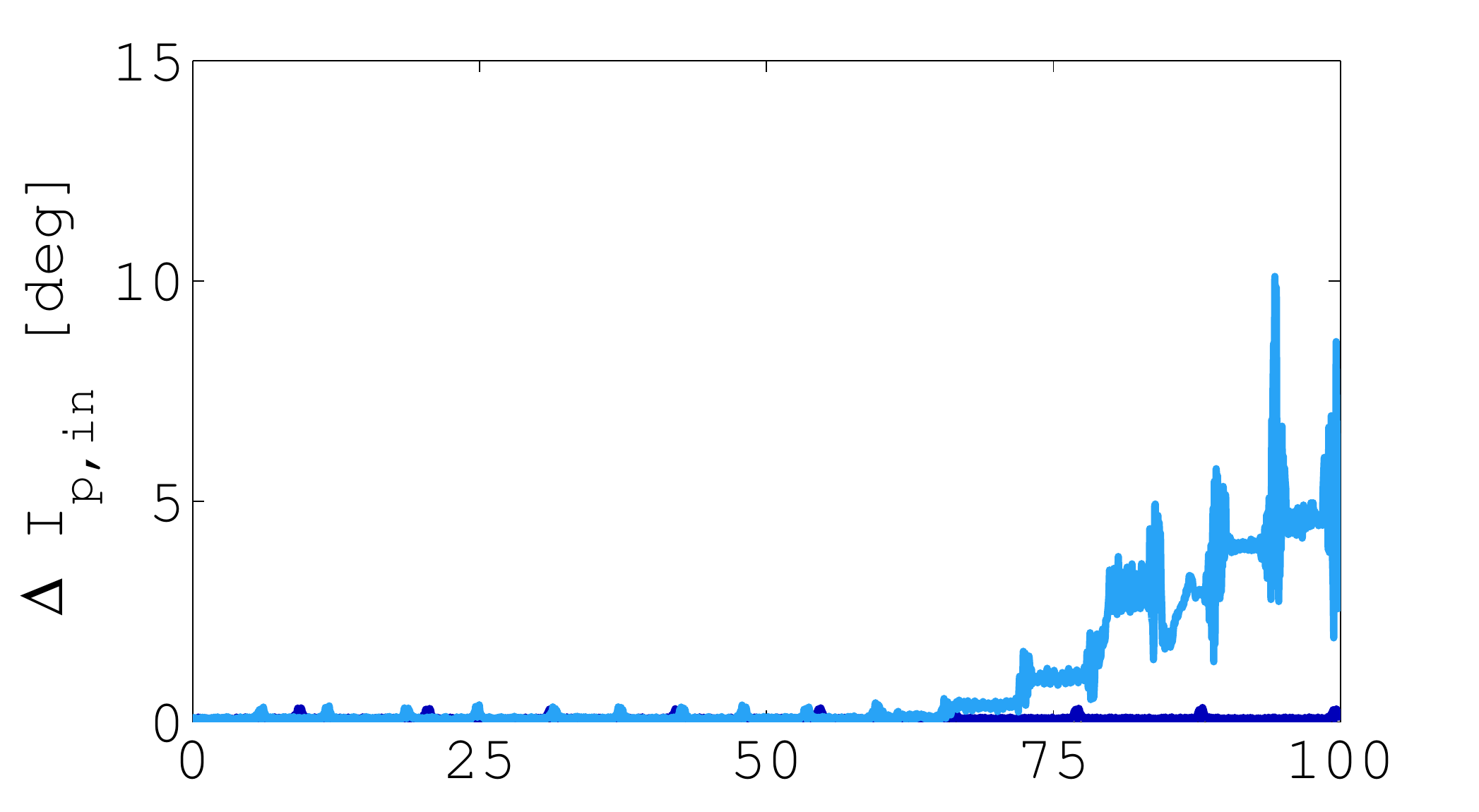}  
	\end{subfigure}
	\vspace{3.5mm}
	\begin{subfigure}[b]{0.49\textwidth}
		\subcaption{{\large$a_{\rm p} = 4 $ AU}}
		\includegraphics[width=\textwidth]{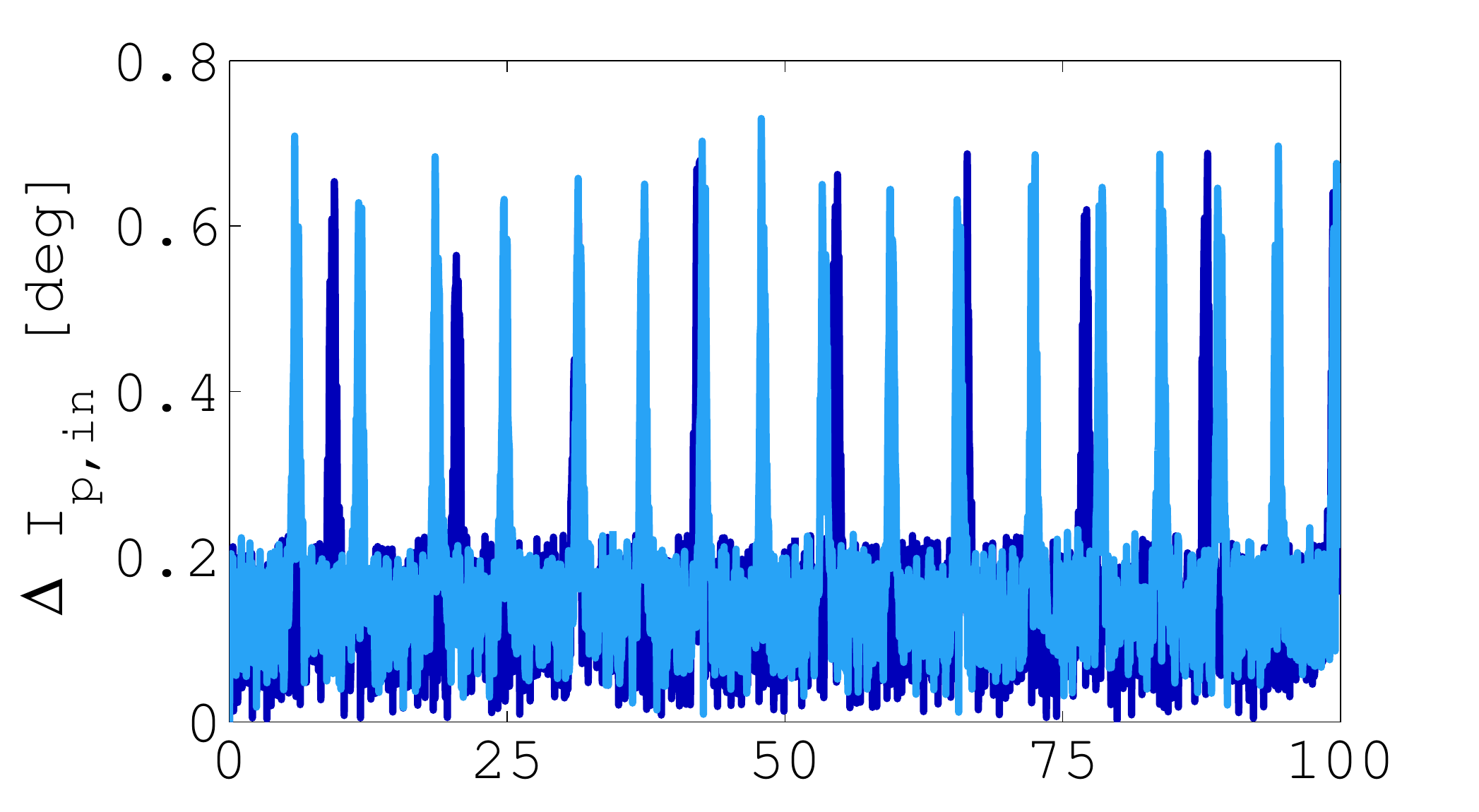}  
	\end{subfigure}
	\begin{subfigure}[b]{0.49\textwidth}
		\subcaption{{\large$a_{\rm p} = 7 $ AU}}
		\includegraphics[width=\textwidth]{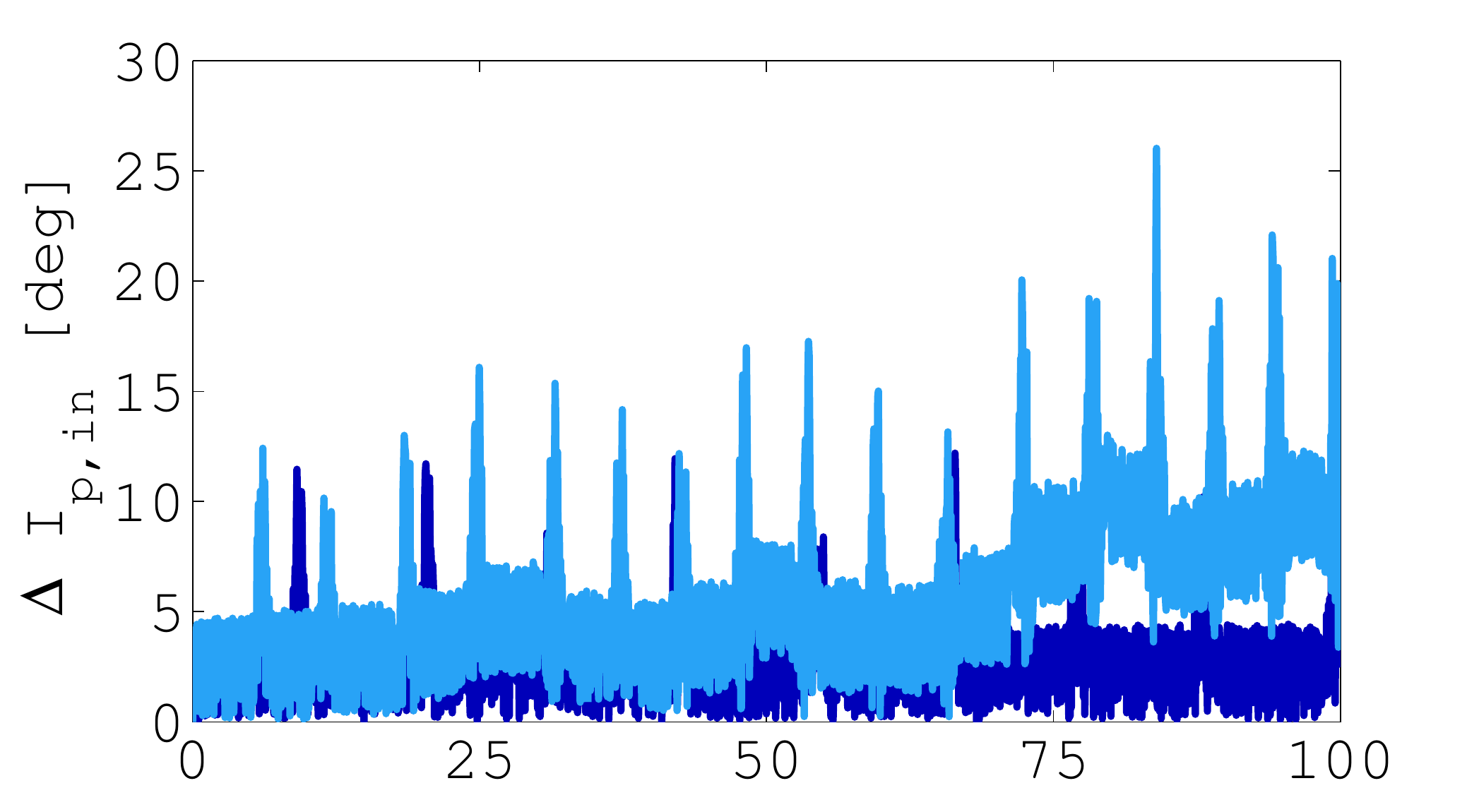}
	\end{subfigure}
	\vspace{3.5mm}
	\begin{subfigure}[b]{0.49\textwidth}
		\subcaption{{\large$a_{\rm p} = 8 $ AU}}
		\includegraphics[width=\textwidth]{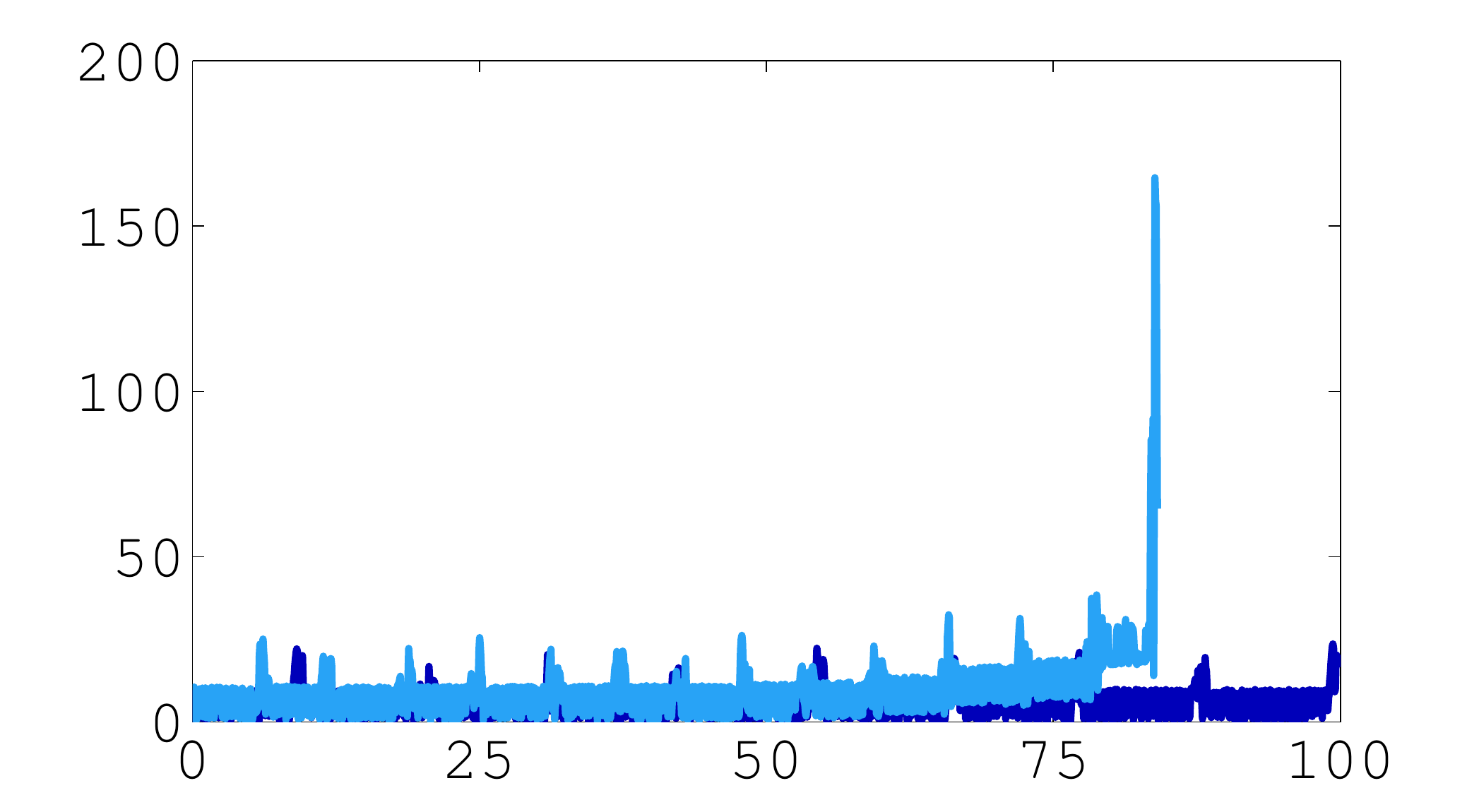}  
	\end{subfigure}
	\begin{subfigure}[b]{0.49\textwidth}
		\subcaption{{\large$a_{\rm p} = 9 $ AU}}
		\includegraphics[width=\textwidth]{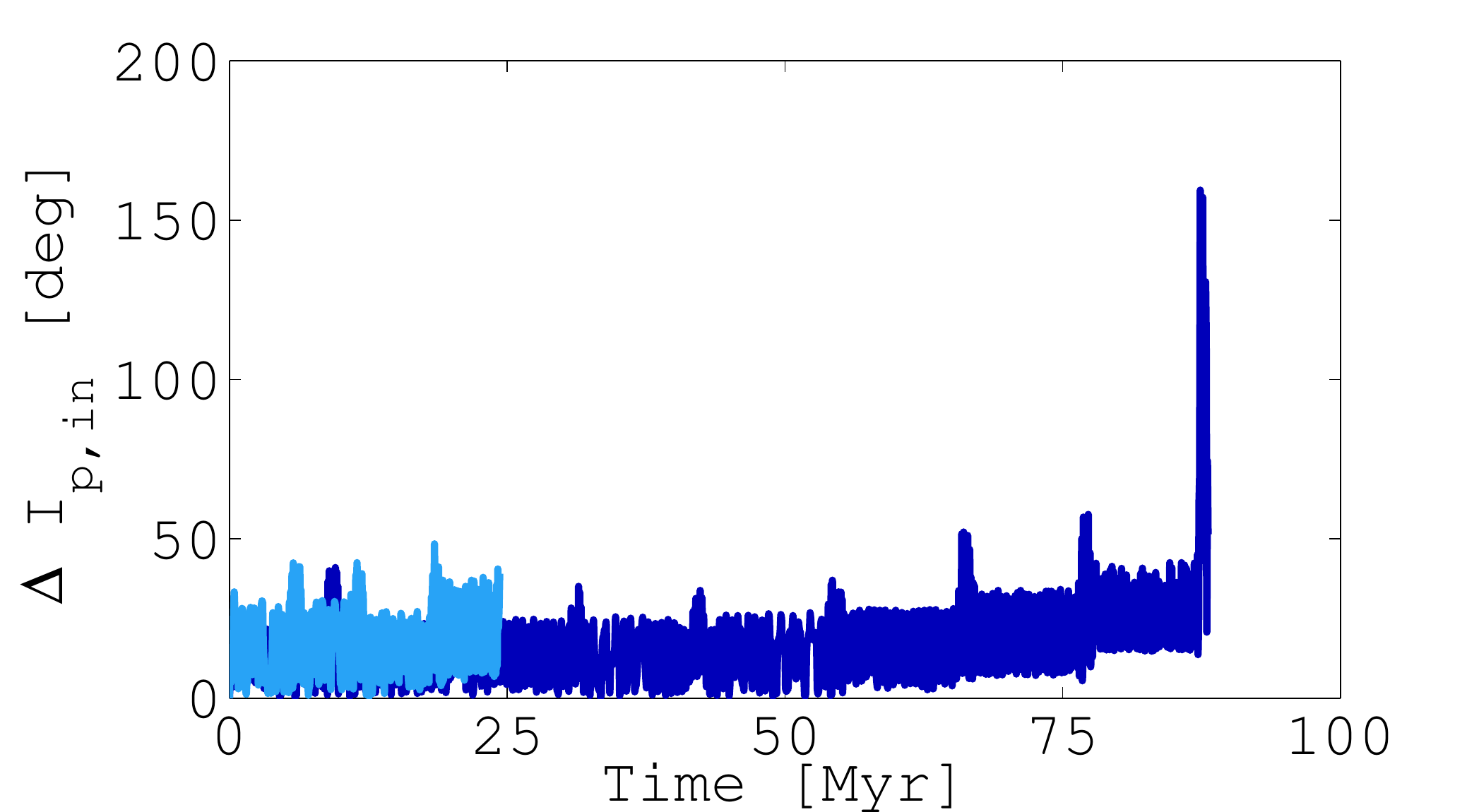}
	\end{subfigure}
	\begin{subfigure}[b]{0.49\textwidth}
		\subcaption{{\large Inner binary}}
		\includegraphics[width=\textwidth]{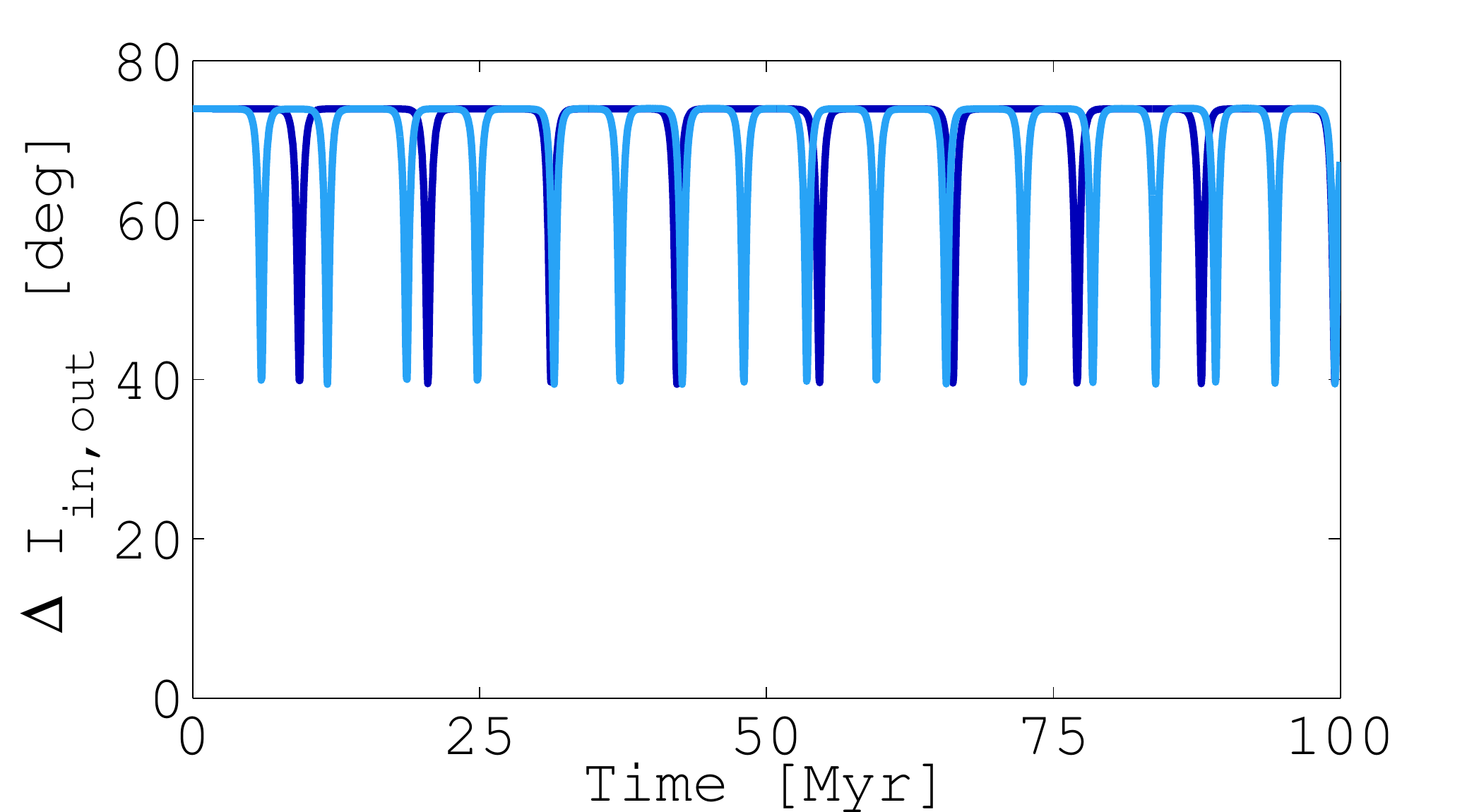}  
	\end{subfigure}
	\caption{ First seven figures: evolution over 100 Myr of the mutual inclination between the planet and the inner binary in a primordial triple star system, where $a_{\rm in} = 0.4827$ AU ($P_{\rm in} = 100$ d), $a_{\rm out} = 100$ AU ($P_{\rm out} = 2.52  \times 10^5$ d) and $e_{\rm out} =$ 0 (dark blue) and 0.25 (light blue) and the inner binary starts on a circular orbit. The planet is massless and starting on an orbit that is circular and coplanar with the inner binary. Each subplot shows a different value of $a_{\rm p}$ (see Table.~\ref{tab:PrimordialParams} for all the simulation parameters). Note the changing scale on the y-axis as a function for different $a_{\rm p}$. Bottom right figure: evolution of the mutual inclination between the inner and outer binaries for $e_{\rm out} =$ 0 (dark blue) and 0.25 (light blue).
}\label{fig:example_DeltaI}  
\end{center}  
\end{figure*}    

Now define
\begin{eqnarray}
\label{eq:alpha}
\alpha &\equiv& \frac{M_{\rm p}}{M_3} \frac{a_{\rm out}^3 (1-e_{\rm out}^2)^{3/2} }{a_{\rm p}^3 (1-e_{\rm p}^2)^{3/2}},\\
\label{eq:beta}
\beta &\equiv& \frac{(M_1+M_2)^2 M_{\rm p}}{M_1 M_2 (M_1+M_2+M_{\rm p})}  \left(\frac{a_{\rm p}}{a_{\rm in}} \right)^2, 
\end{eqnarray}
which measure the numerical dominance of the different terms.  For typical circumbinary planets, $\beta \ll 1$ because $M_{\rm p} < M_2 (a_{\rm p}/a_{\rm in})^2$.  In that case, $\langle \mathcal F_{\rm p,out} \rangle $ is negligible compared with $\langle \mathcal F_{\rm in,p} \rangle$ and $\langle \mathcal F_{\rm in,out} \rangle$, and we can find a relatively simple closed-form solution of the eccentricity maximum of the binary.  This is accomplished by making the substitution
\begin{equation}
\sin^2  \Delta I_{\rm in,out}  = 1-\cos^2  \Delta I_{\rm in,out}  = 1- \cos^2  \Delta I_{\rm in,out,init} (1-e_{\rm in}^2)^{-1},
\end{equation}
due the initial condition $e_{\rm in,init}=0$ and the conservation of $H'$. 
The result is an implicit equation of $e_{\rm in}$ versus $\omega_{\rm in}$.  The maximum eccentricity occurs at $\omega_{\rm in}=\pm \pi/2$, which we insert and manipulate 
 to yield:
%-------------------------------
\begin{equation}
e_{\rm in,max} = \left(1 - \frac{5}{3-\alpha} \cos^2  \Delta I_{\rm in,out,init}\right)^{1/2}.
\label{eq:e_max_supressed_binary}
\end{equation}
%------------------------------
This expression recovers Eq.~\ref{eq:kozai_binary_ecc} in the limit $\alpha \rightarrow 0$, and shows that the maximum value of eccentricity cycles declines with increasing planet mass until no eccentricity is excited for masses above:
\begin{equation}
M_{\rm p,crit} = 3 M_3 \frac{a_{\rm p}^3 (1-e_{\rm p}^2)^{3/2}}{a_{\rm out}^3 (1-e_{\rm out}^2)^{3/2} }.
\label{eq:Mp_crit}
\end{equation}

Alternately, for known CBP systems, we may view this as a limit on the properties of a tertiary that can cause the binary to shrink by Kozai cycles with tidal friction: 
\begin{equation}
%\Big{(} {\frac{ M_3 }{a_{\rm out}^3 (1-e_{\rm out}^2)^{3/2} } \Big{)}_{\rm crit} = \frac{M_{\rm p}}{3 a_{\rm p}^3 (1-e_{\rm p}^2)^{3/2} }.
\left(\frac{ M_3 }{a_{\rm out}^3 (1-e_{\rm out}^2)^{3/2}}\right)_{\rm crit}
=
 \frac{M_{\rm p}}{3 a_{\rm p}^3 (1-e_{\rm p}^2)^{3/2} },
\end{equation}
where the left and right hand sides quantify the strength of the tidal force from the tertiary star and planet, respectively. Kozai is suppressed on the inner binary when the left hand side is larger.
The only known CBP that has a third distant companion is PH-1/Kepler-64; this companion is itself a binary (Ba-Bb) of total mass $\sim 1.5M_\odot$.  The planet's mass is not known, but its radius of $\sim 6 R_\oplus$ suggests its mass is at least of order Neptune's, i.e. $15 M_\oplus$.   The critical value of the planet's tidal parameter, the right hand side of the last equation, is therefore $6 \times 10^{-5} M_{\odot} \rm{AU}^{-3}$.  On the other hand, the projected separation of Ba-Bb to the host binary is $\sim1000$~AU.  Supposing its eccentricity is very high and so its semi-major axis is only $a_{\rm out}=500$~AU, then the critical value of the parameters is reached only for $e_{\rm out} = 0.998$.  This is too high to allow the planet to remain dynamically stable.  Hence we essentially rule out the possibility that this companion will excite Kozai cycles in the planet-hosting binary, because the planet precesses its host too fast to allow this to happen.

%====================
%Section
\section{N-body simulations of a set of example systems}
\label{sec:example}
%====================

In this section we use n-body simulations to determine regions where planets may be able to form and survive in evolving triple star systems, and whether or not the binary will be allowed to shrink under KCTF. We take a stable orbit to be a necessary but probably not sufficient condition for planetary formation. We are therefore considering the most optimistic scenario possible. Additional effects are considered afterwards in Sect.~\ref{sec:additional_effects}. The entire argument is summarised in Sect.~\ref{sec:general_argument}.
%====================
%Subsubsection 
\subsection{A primordial inner binary}
\label{sec:example_primordial}
%====================

%---------------------------------------------------------------------
\begin{table}
\caption{Parameters for the primordial triple star system and circumbinary planet} %title of the table
\centering % centering table
\begin{tabular}{l | l | l | } % creating eight columns
\hline\hline %inserting double-line
inner binary (one set of parameters) & \\
$M_1$ & 1 $M_{\odot}$\\
$M_2$ & 0.5 $M_{\odot}$\\
$a_{\rm in}$ & 0.4827 AU\\
$P_{\rm in}$ & 100 d\\
$e_{\rm in}$ & 0\\
\hline % inserts single-line
circumbinary planet (18 sets of parameters) & \\
$M_{\rm p}$ & 0\\
$a_{\rm in}$ & 1.5 --- 10 AU\\
$P_{\rm in}$ & 548 --- 9430 d\\
$e_{\rm p}$ & 0\\
$\Delta I_{\rm p,in}$ & 0\\
\hline % inserts single-line
outer binary (two sets of parameters)& \\
$M_3$ & 0.6 $M_{\odot}$\\
$a_{\rm out}$ & 100 AU\\
$P_{\rm out}$ & $2.52 \times 10^5$ d\\
$e_{\rm out}$ & 0, 0.25\\
$\Delta I_{\rm in,out}$ & $74^{\circ}$\\
\hline % inserts single-line

\end{tabular}
\label{tab:PrimordialParams}
\end{table}
%--------------------------------------------------

For our first suite of n-body simulations we constructed an example primordial triple star system, with an interior circumbinary planet, using the parameters listed in Table~\ref{tab:PrimordialParams}. For the inner binary we tested a single set of parameters, where the orbit was arbitrarily chosen at 100 d and the stellar masses were the average values for the circumbinary systems discovered by {\it Kepler} ($M_1 = 1M_{\odot}$ and $M_2 = 0.5M_{\odot}$). The inner eccentricity was set to zero initially, although it rose dramatically during Kozai cycles.

The tertiary star was given a mass $M_3 = 0.6M_{\odot}$, corresponding to the median observed outer mass ratio $q_{\rm out} = 0.4$, and placed at a typical separation of 100 AU \citep{tokovinin08}. The mutual inclination $\Delta I_{\rm in,out} = 74^{\circ}$ was chosen such that the inner binary would ultimately shrink to a 5 d period, according to Eq.~\ref{eq:final_binary_a}, where $e_{\rm in,max} = 0.93$ (Eq.~\ref{eq:kozai_binary_ecc}). We tested two different tertiary star orbits where we only changed $e_{\rm out}$ = 0 and 0.25, to test the effect it has on planetary stability. We also tested $e_{\rm out} = 0.5$ but found that almost all systems were unstable in this configuration so we have omitted these results for simplicity.  The requirement that $e_{\rm out}<0.5$ for stability is restrictive because observations show this to be a typical outer eccentricity \citep{tokovinin08}.

Orbiting around the inner binary we placed a massless planet on a circular and coplanar orbit, like in Fig.~\ref{fig:geometry}. The behaviour and stability of the planet are largely functions of the $a_{\rm p}$, as this determines the relative perturbing strengths from the inner and outer binaries. For the simulations we tested 18 different values of $a_{\rm p}$ between 1.5 and 10 AU, in steps of 0.5 AU.

The range of semi-major axes was chosen based on the relative Kozai and precession timescales on the planet. We plot these competing timescales in Fig.~\ref{fig:competing_timescales} as a function of $a_{\rm p}$. The Kozai timescale in the planet is a monotonic decreasing function of $a_{\rm p}$ (Eq.~\ref{eq:tau_kozai_planet})\footnote{To apply this equation to a triple star system one replaces $M_1$ by $M_1+M_2$ and $M_2$ by $M_3$.}. We plot $\tau_{\rm Kozai,p}$ the two values of $e_{\rm out}$. On this logarithmic scale $e_{\rm out}$ does not have a discernible affect on $\tau_{\rm Kozai,p}$. The circumbinary precession timescale (Eq.~\ref{eq:tau_prec}), on the other hand, increases with $a_{\rm p}$. For $P_{\rm in} = 100$ d there is a turnover between the two timescales at approximately $a_{\rm p} = 7$ AU. For farther out planetary orbits we expect the planet to undergo Kozai cycles and obtain an eccentricity $e_{\rm p, max} = 0.93$, leading to ejection. For smaller $a_{\rm p}$, we expect Kozai to be suppressed and the planet to remain stable, as long as it is beyond the circumbinary stability limit in Eq.~\ref{eq:stability_limit_CB}.

Using the REBOUND code we ran n-body integrations for the total of 36 simulations. These were run over 100 Myr, which lets us cover almost 100 Kozai timescales ($\tau_{\rm Kozai, in} = 1.17 \times 10^6$ yr from Eq.~\ref{eq:tau_kozai_binary}). When the tertiary star is eccentric there is also a longer octupole timescale,

\begin{equation}
\tau_{\rm oct,in} \sim \left(\frac{a_{\rm out}(1-e_{\rm out}^2)}{e_{\rm out}a_{\rm in}}\right)^{1/2}\tau_{\rm Kozai,in}
\end{equation}
\citep{antognini15}. In our example $\tau_{\rm oct,in} = 3.27 \times 10^7$ yr, and hence we are covering a couple of these timescales, but may be missing some dynamical effects that develop over many timescales. In our simulations a planet was considered unstable if it were ejected from the system.

In Fig.~\ref{fig:example_eccentricity} we plot $e_{\rm p}$ over time for a selection of simulations: $a_{\rm p} =$ 1.5, 2, 3.5, 4, 7, 8 and 9 AU. In each plot we show the result for $e_{\rm out}$ = 0 (dark blue) and 0.25 (light blue). For reference, in the bottom right figure we show the inner binary over time. The maximum $e_{\rm in}$ is negligibly different in the two cases: $e_{\rm in,max} = 0.9239$ for $e_{\rm out} = 0$ and $e_{\rm in,max} = 0.9241$ for $e_{\rm out} = 0.25$ but the Kozai modulation period changes significantly: $\sim 10$ yr for a circular tertiary and $\sim 5.5$ yr for an eccentric tertiary.

For $a_{\rm p} = 1.5$ AU neither configuration is stable, as in each case the planet is ejected as soon as the inner binary reaches its eccentricity maximum for the first time. When the planet is a little farther from the inner binary, at 2 AU, and $e_{\rm out = 0}$, its eccentricity rises to 0.1 periodically, coinciding with when the inner binary undergoes its Kozai modulation, but the planet nevertheless remains stable over 100 Myr. This roughly calculated inner stability limit coincides well with Eq.~\ref{eq:stability_limit_CB} from \citet{holman99}, where $a_{\rm crit,CB} = 1.94$ AU for $e_{\rm in} = e_{\rm in,max} = 0.94$. When $e_{\rm out}$ is increased to 0.25 the planet is no longer stable at 2 AU. In fact the shortest-period planet that was stable over 100 Myr for $e_{\rm out} = 0.25$ was $a_{\rm p} = 3.5$ AU, although even in this configuration the planet obtained significant eccentricity that may have led to instability on a longer timescale. For $a_{\rm p} = 4$ AU there is no large eccentricity variation and the results $e_{\rm out}$ = 0 and 0.25 are almost identical. This may be considered the inner stability limit for $e_{\rm out}$ = 0.25.

The inner stability limits for $e_{\rm out}$ = 0 (2 AU) and 0.25 (4 AU) are significantly different. This is despite the maximum inner binary eccentricity being practically the same in the two cases. Qualitatively, the difference in stability limits likely due to an increased eccentricity obtained by the planet, due to the eccentric third star, pushing the planet closer to the inner binary and causing instability.

 For $e_{\rm out} = 0.25$ the planet remains stable for $a_{\rm p} < 8$ AU. For a circular tertiary the planet remains stable a little father out, before reaching instability at 9 AU. This is near where $\tau_{\rm Kozai,p}$ was seen to become shorter than $\tau_{\rm prec,p}$ in Fig.~\ref{fig:competing_timescales}. These outer stability limits are on a significantly shorter-period than predicted by the circumprimary stability limit in Eq.~\ref{eq:stability_limit_CP} from \citet{holman99}: 35.54 AU for $e_{\rm out} = 0$ and 24.54 AU for $e_{\rm out} = 0.25$. This is because \citet{holman99} was calculated for a circular and coplanar planet, and not one that may potentially undergo Kozai cycles.

In Fig.~\ref{fig:example_DeltaI} we plot $\Delta I_{\rm p,in}$ for the same simulations, and $\Delta I_{\rm in,out}$ in the bottom right figure. An ejection is often preceded by a large rise in the mutual inclination. For stable systems, there is a trend for the planets closer to the inner binary to remain close to coplanarity, whilst the planets farther away obtain at least a few degrees of mutual inclination. This means that even if Kozai cycles are suppressed, the tertiary star still has a perturbing effect.

%Stable orbits were seen for planets orbiting between 2 and 7.8 AU, which we denote using solid red vertical lines in Fig.~\ref{fig:competing_timescales}. 

%The work of \citet{holman99}, which neglects $e_{\rm p}$, calculates an inner limit at ${\bf 1.95 AU}$ (Eq.~\ref{eq:stability_limit_CB}) and an outer limit at 24 AU (Eq.~\ref{eq:stability_limit_CP}). There is a close match with our numerical work at the inner limit because $e_{\rm p}=0$ here due to Kozai suppression.  The outer limit of \citet{holman99} is a large overestimate because it does not take into account an induced Kozai cycle on the planet.

%It may be possible to have a planet stably undergo Kozai in a wider triple system. However since 70 AU is the median observed tertiary semi-major axis, at least half of the configurations are probably incompatible with a Kozai planet. 

%Under this basic analysis, it is thought that planets may be able to form within a region of several AU width, relatively far out in the disc. A reduction of the ratio $a_{\rm out}/a_{\rm in, init}$ by a factor of $\sim4$ would make this region completely vanish.

%====================
%Subsubsection 
%====================

\begin{figure}  
\begin{center}  
\includegraphics[width=0.49\textwidth]{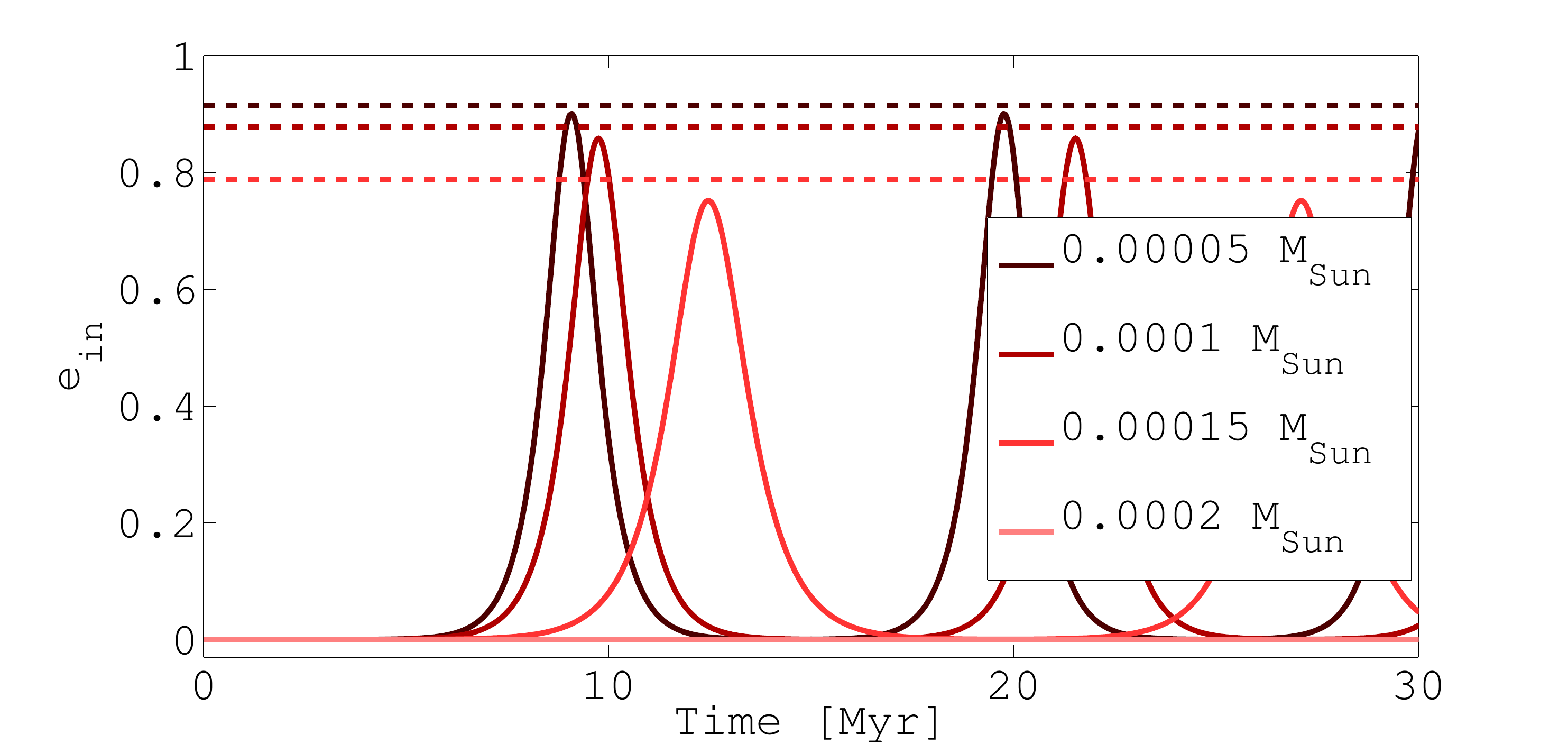} 
\caption{Inner binary eccentricity over time for $a_{\rm p} = 5$ AU, $e_{\rm out = 0}$ and a varied $M_{\rm p}$. Solid lines are from numerical integration and horizontal dashed lines are the eccentricity maxima calculated using Eq.~\ref{eq:e_max_supressed_binary}. For $m_{\rm p} = 0.0002 M_{\odot}$ both the numerical and analytic results show $e_{\rm in,max} = 0$, i.e. a line across the bottom horizontal axis.}
\label{fig:planet_mass_test}  
\end{center}  
\end{figure} 

\begin{figure}  
\captionsetup[subfigure]{labelformat=empty}
\begin{center}  
	\begin{subfigure}[b]{0.49\textwidth}
		\subcaption{{\large $a_{\rm p} = 4 $ AU, $e_{\rm out} = 0$}}
		\includegraphics[width=\textwidth]{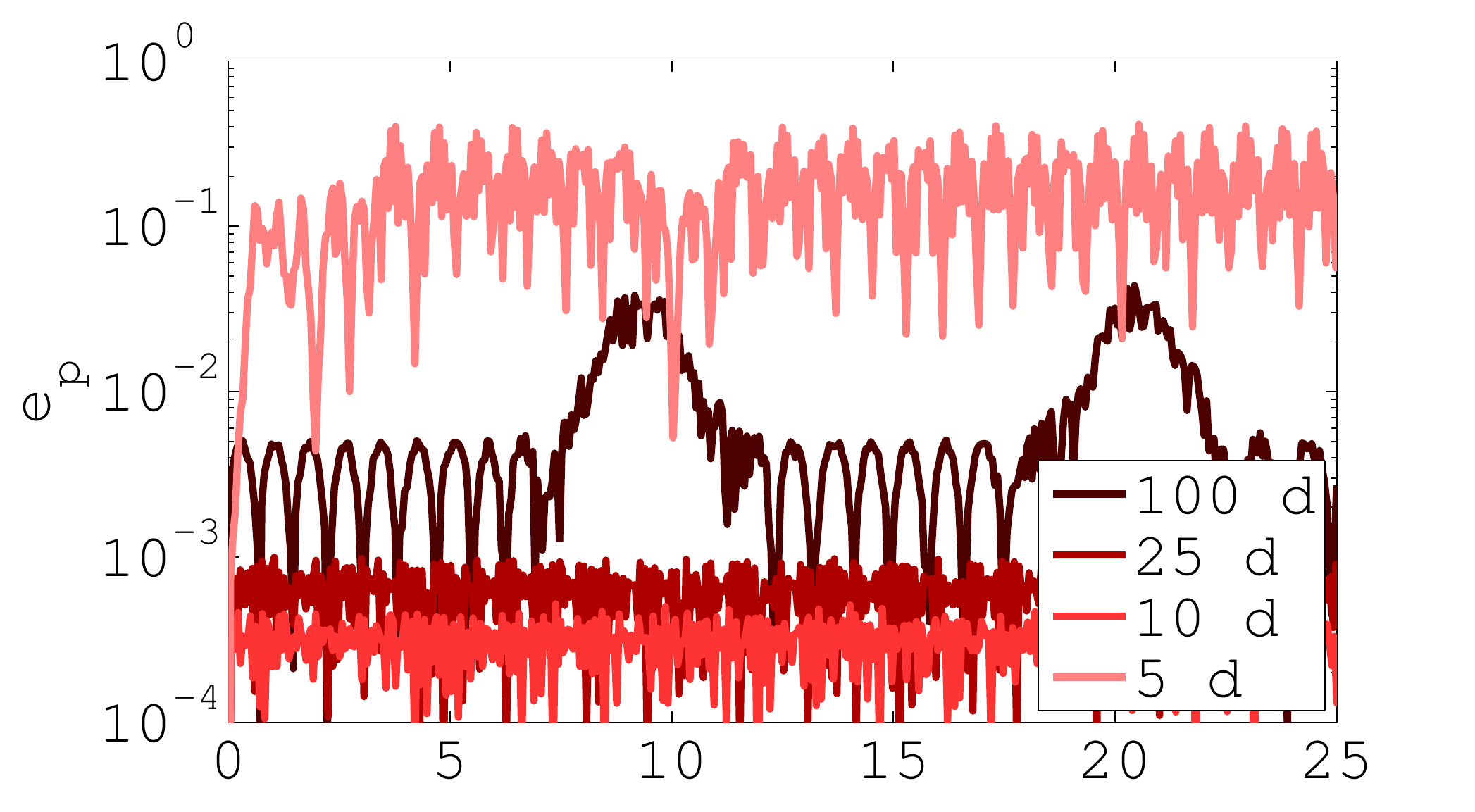}  
	\vspace{3.5mm}
	\end{subfigure}
	\begin{subfigure}[b]{0.49\textwidth}
		\subcaption{{\large$a_{\rm p} = 4 $ AU, $e_{\rm out} = 0.25$}}
		\includegraphics[width=\textwidth]{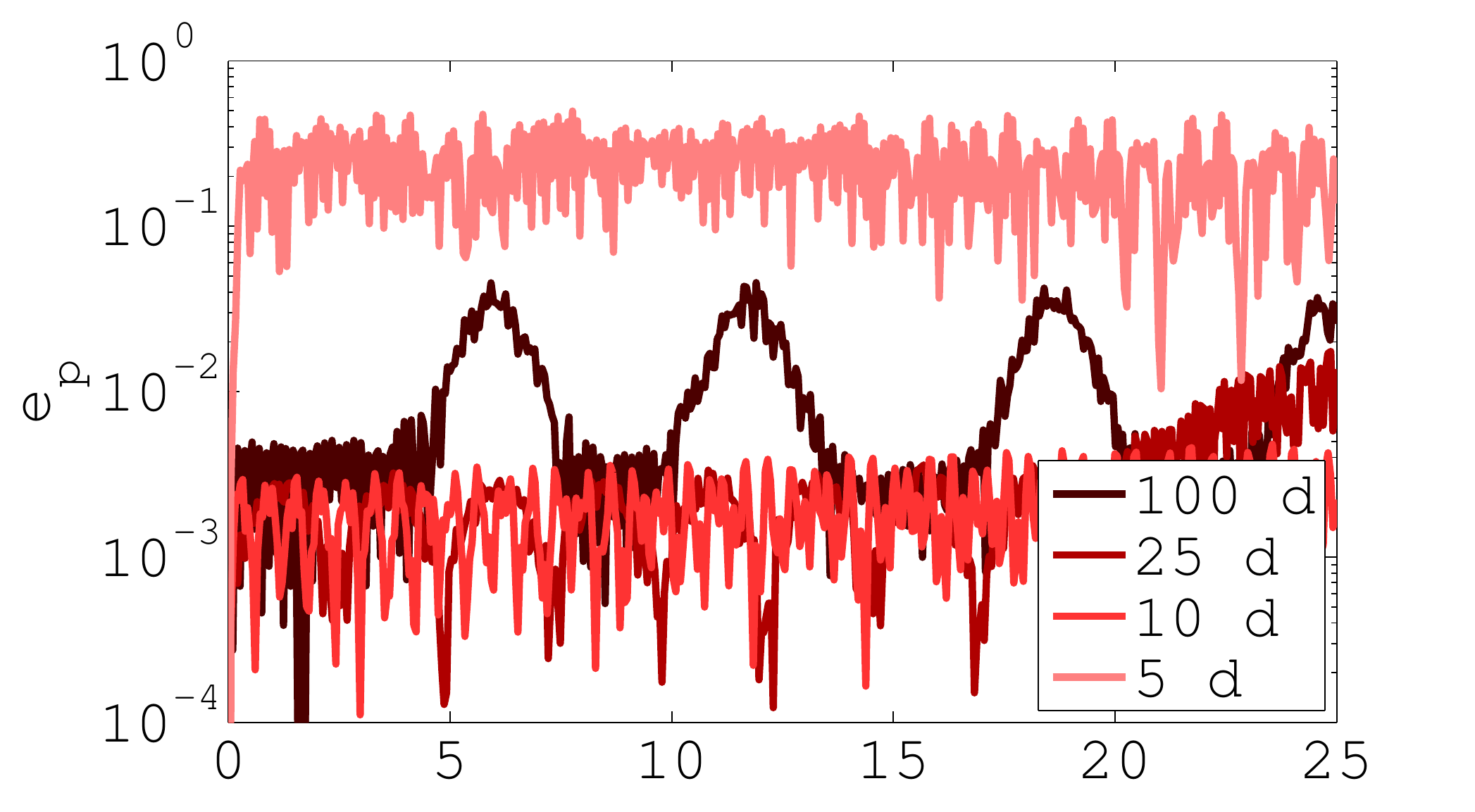}  
	\vspace{3.5mm}
	\end{subfigure}	
	\begin{subfigure}[b]{0.49\textwidth}
		\subcaption{{\large$a_{\rm p} = 5 $ AU, $e_{\rm out} = 0$}}
		\includegraphics[width=\textwidth]{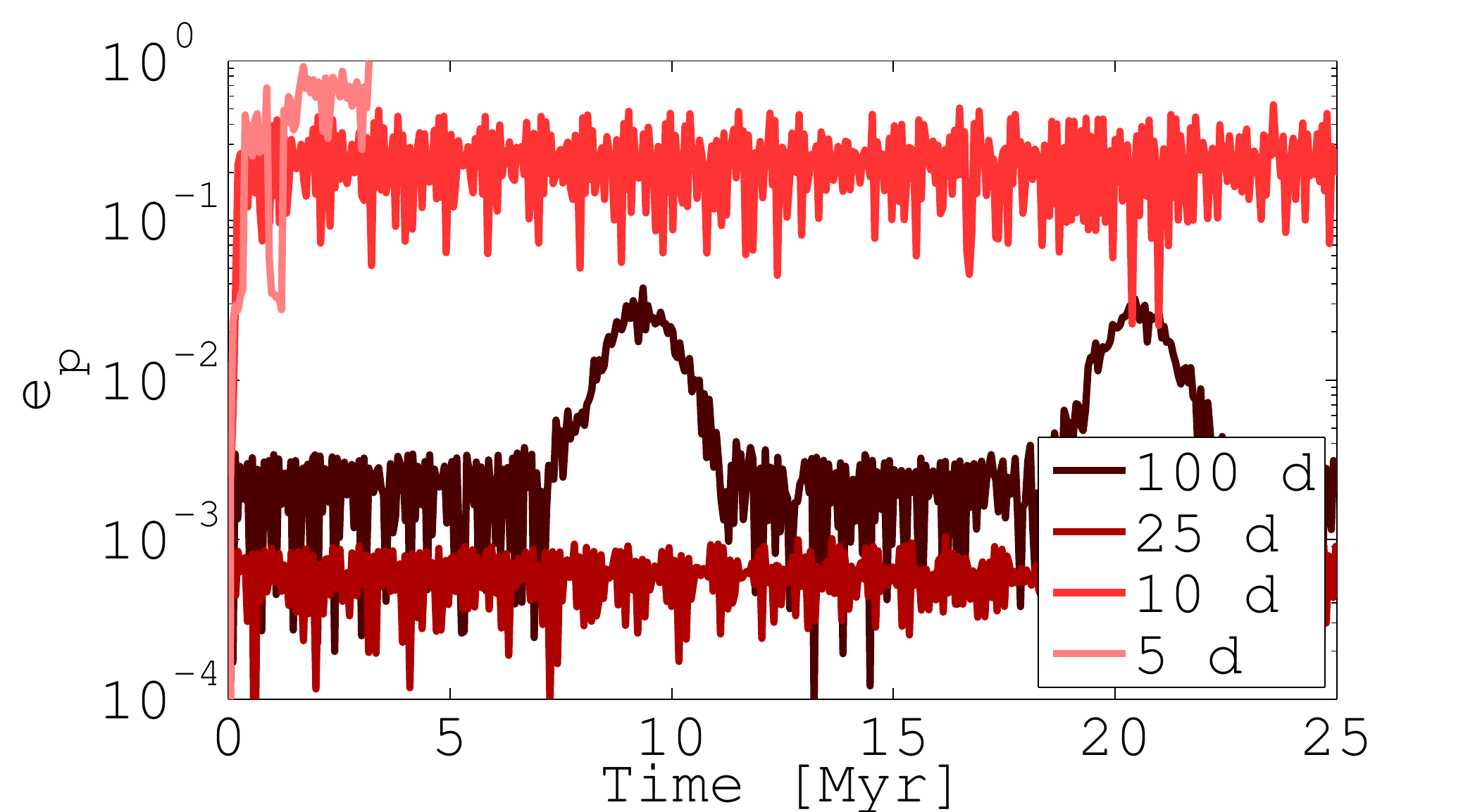}  
	\end{subfigure}
	\caption{ Evolution of the planet's eccentricity over 25 Myr around an inner binary of period 100, 25, 10 and 5 d, in the presence of an outer tertiary star at 100 AU. In the top figure $a_{\rm p} = 4$ AU and $e_{\rm out} = 0$. In the middle figure $a_{\rm p} = 4$ AU and $e_{\rm out} = 0.25$. In the bottom figure $a_{\rm p} = 5$ AU and $e_{\rm out} = 0$.
}\label{fig:shrinkage_eccentricity}  
\end{center}  
\end{figure} 

\begin{figure}  
\captionsetup[subfigure]{labelformat=empty}
\begin{center}  
	\begin{subfigure}[b]{0.49\textwidth}
		\subcaption{{\large$a_{\rm p} = 4 $ AU, $e_{\rm out} = 0$}}
		\includegraphics[width=\textwidth]{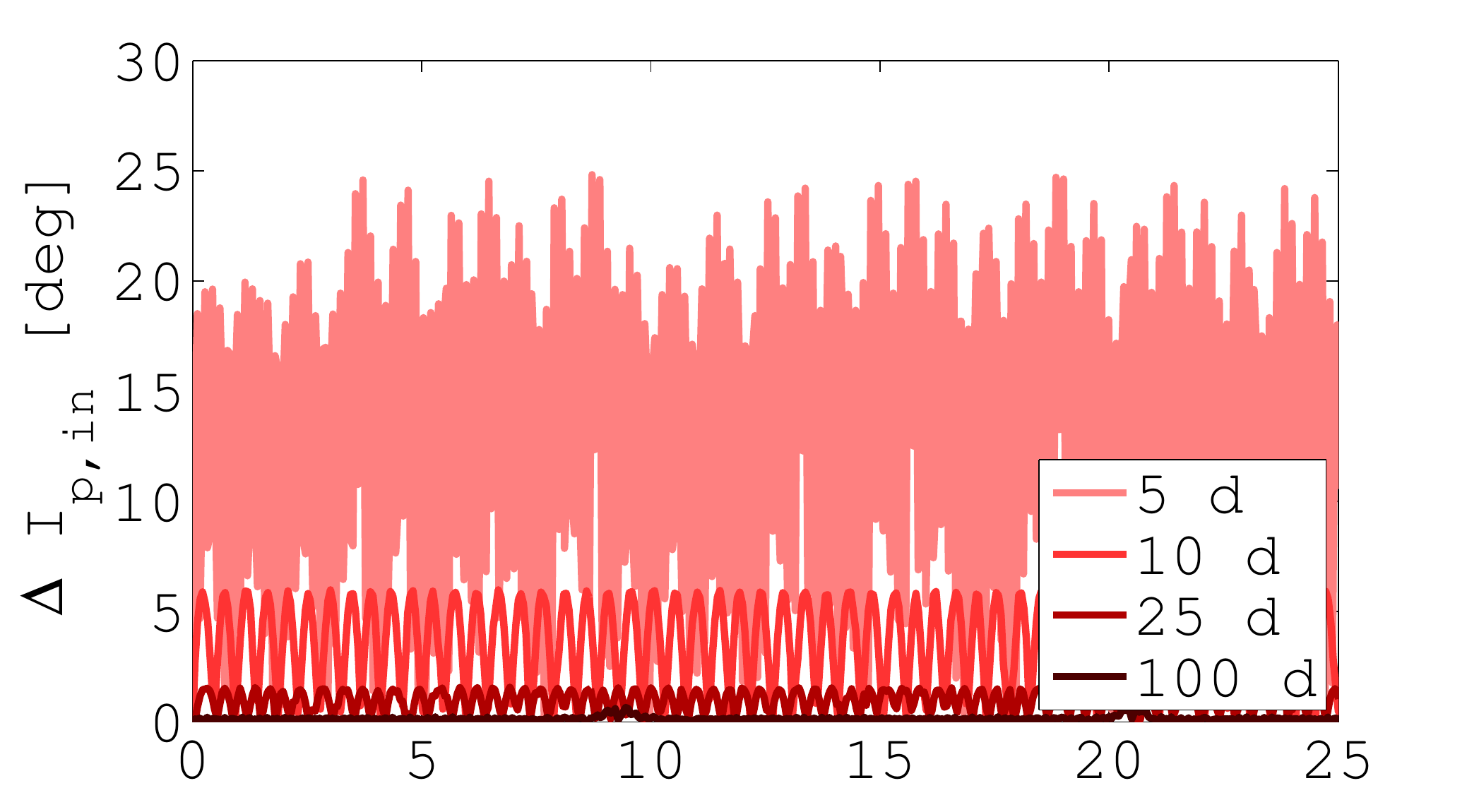}  
	\vspace{3.5mm}
	\end{subfigure}
	\begin{subfigure}[b]{0.49\textwidth}
		\subcaption{{\large$a_{\rm p} = 4 $ AU, $e_{\rm out} = 0.25$}}
		\includegraphics[width=\textwidth]{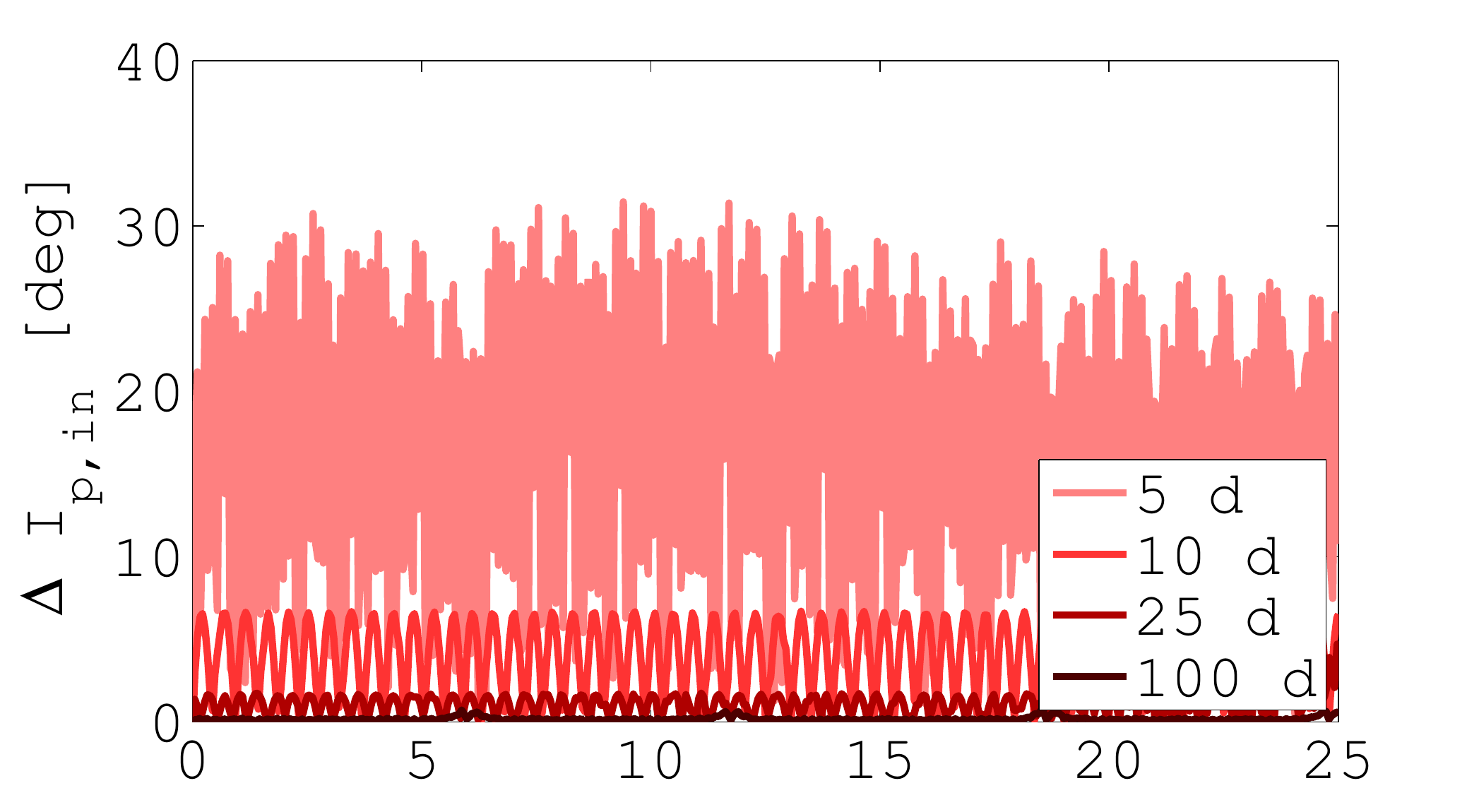}  
	\vspace{3.5mm}
	\end{subfigure}	
	\begin{subfigure}[b]{0.49\textwidth}
		\subcaption{{\large$a_{\rm p} = 5 $ AU, $e_{\rm out} = 0$}}
		\includegraphics[width=\textwidth]{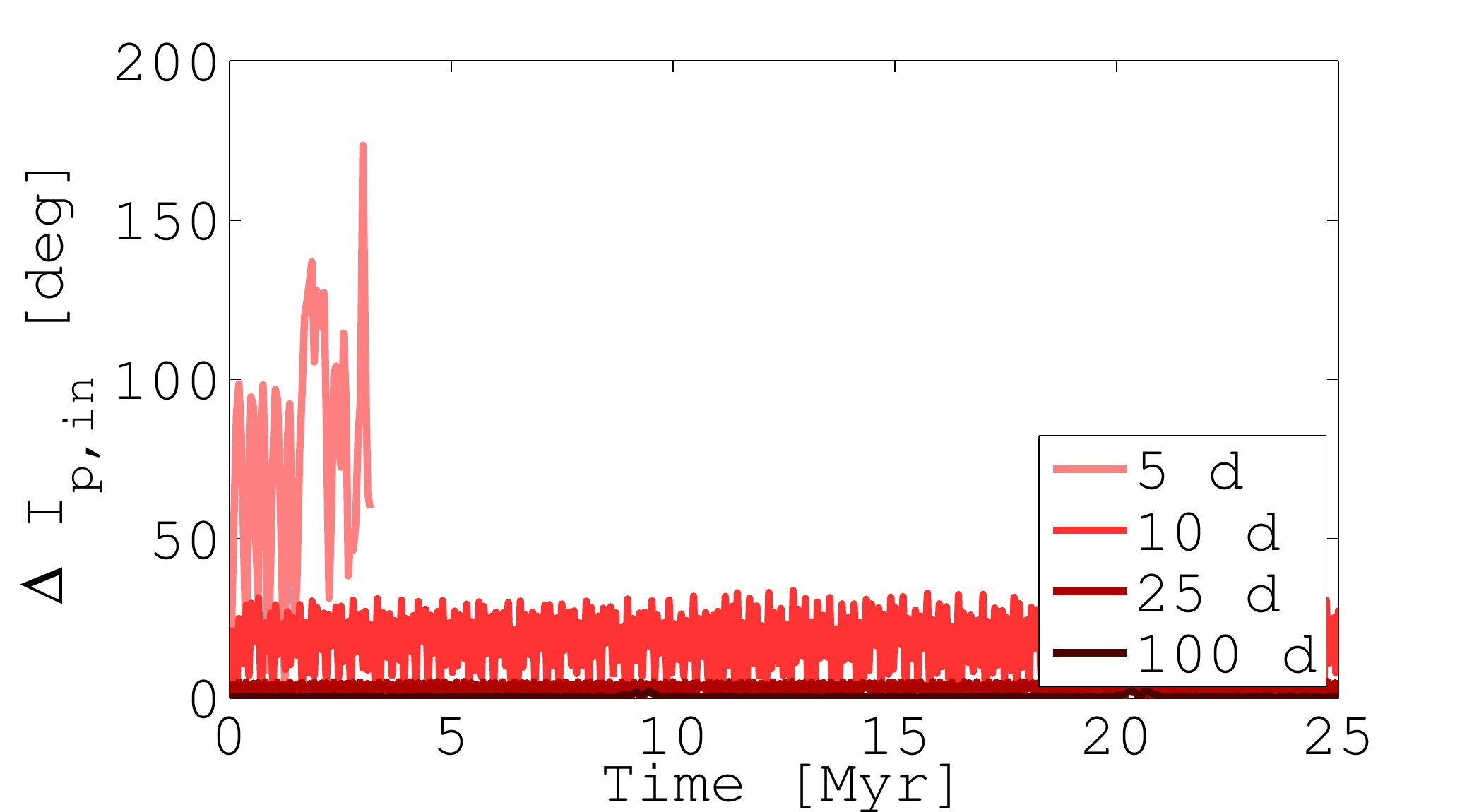}  
	\end{subfigure}
	
	\caption{ Evolution of the mutual inclination between the planet and the inner binary, starting coplanar, over 25 Myr around an inner binary of period 100, 25, 10 and 5 d, in the presence of an outer tertiary star at 100 AU, for the same simulations as in Fig.~\ref{fig:shrinkage_eccentricity}. In the top figure $a_{\rm p} = 4$ AU and $e_{\rm out} = 0$. In the middle figure $a_{\rm p} = 4$ AU and $e_{\rm out} = 0.25$. In the bottom figure $a_{\rm p} = 5$ AU and $e_{\rm out} = 0$.
}\label{fig:shrinkage_DeltaI}  
\end{center}  
\end{figure}

\begin{figure}  
\begin{center}  
\includegraphics[width=0.49\textwidth]{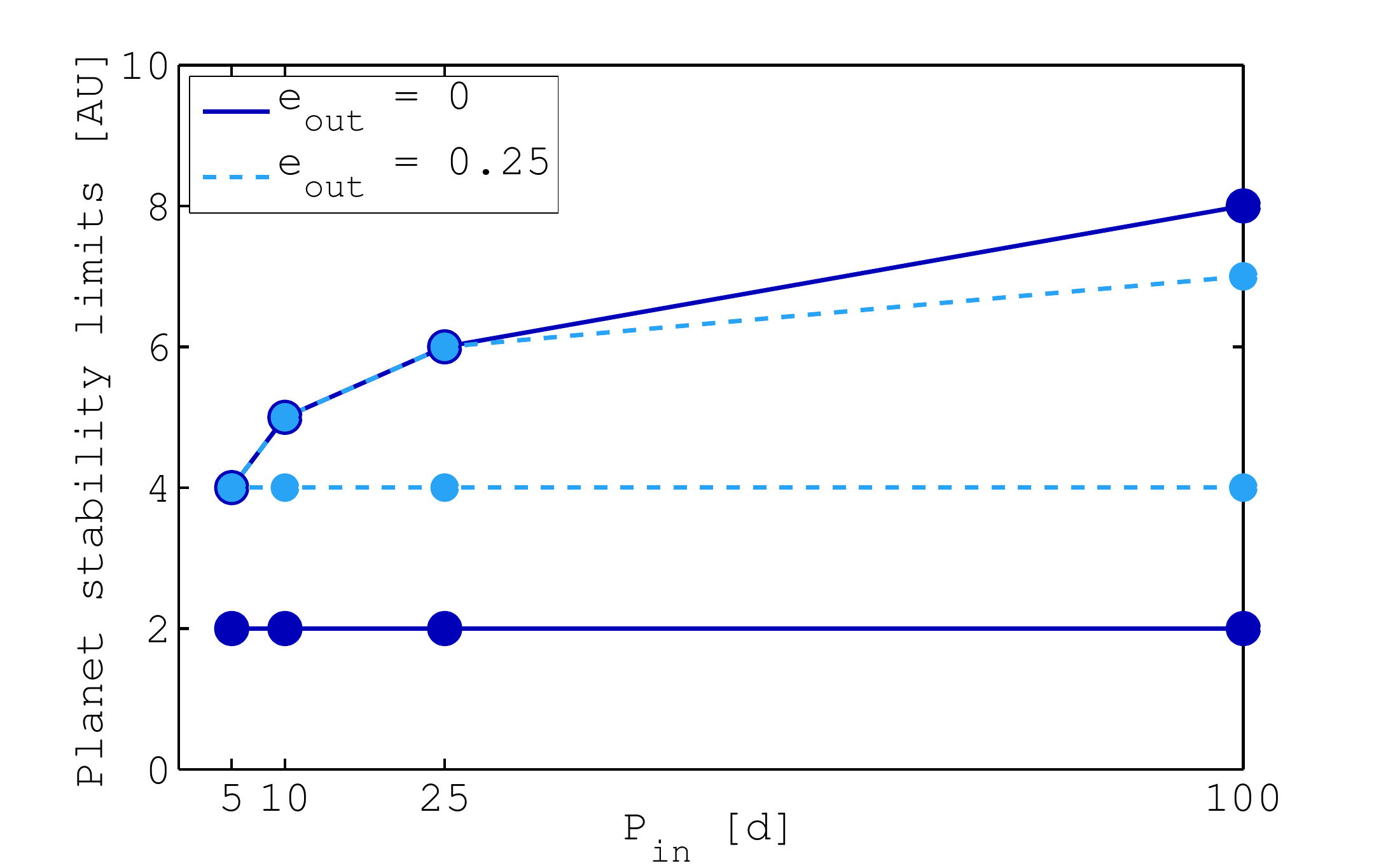} 
\caption{Numerically determined inner and outer stability limits for the planet as a function of $P_{\rm in}$ for $e_{\rm out}$ = 0 (dark blue, solid) and 0.25 (light blue, dashed). The inner stability limit is defined from the test of the primordial binary in Sect.~\ref{sec:example_primordial}, and this does not change as the binary shrinks.}
\label{fig:stability_limits}  
\end{center}  
\end{figure}

\subsection{What planets would suppress the KCTF process?}

The next question we have to consider is whether our binary orbit, with its primordial orbit of 100 d, will undergo KCTF without suppression by the planet. In Fig.~\ref{fig:planet_mass_test} we plot $e_{\rm in}$ over time for a planet at 5 AU with different planetary masses, in steps of 0.00005 $M_{\odot}$. In this simulation $e_{\rm out} = 0$. We compare these simulations with the analytic prediction from Eq.~\ref{eq:e_max_supressed_binary}, showing that it is accurate to within roughly 5\%. This small discrepancy may arise from Eq.~\ref{eq:e_max_supressed_binary} being derived under the assumption that $\beta = 0$ in Eq.~\ref{eq:beta}. 

For a massless planet $e_{\rm in,max} = 0.9345$, and the expected final period of the inner binary is 5 d (Eq.~\ref{eq:final_binary_a}). A planet of mass 0.00005 $M_{\odot}$ = 16.65 $M_{\oplus}$ causes a small reduction in $e_{\rm in,max} = 0.9150$. When $m_{\rm p}$ is increased to 0.00015 $M_{\odot}$ = 50 $M_{\oplus}$ the binary eccentricity is limited to 0.7875. In this scenario it is likely that the inner two stars never get close enough for tidal forces to have a noticeable effect over their lifetime, and hence there is likely to be no tidal shrinkage. A further increase in $m_{\rm p}$ to 0.0002 $M_{\odot}$ = 67 $M_{\oplus}$ completely suppresses the Kozai modulation.

For planets at the inner (2 AU) and outer (8 AU) edges of the stability range, the critical planet masses for Kozai suppression from Eq.~\ref{eq:Mp_crit} are 5 $M_{\oplus}$ (0.016 $M_{\rm Jup}$) and 300 $M_{\oplus}$ (0.94 $M_{\rm Jup}$), respectively.

\subsection{A shrinking inner binary}
\label{sec:shrinking_inner_binary}

If the planet mass is sufficiently small such that the inner binary Kozai cycles are not suppressed, one would expect its orbit to shrink from $a_{\rm in,init} = 0.4827$ AU (100 d) to $a_{\rm in,final} = 0.0655$ AU (5 d), according to the approximation in Eq.~\ref{eq:final_binary_a}. As the inner binary shrinks to 5 d the circumbinary precession timescale increases (Eq.~\ref{eq:tau_prec}). We demonstrate this in Fig.~\ref{fig:competing_timescales} by plotting $\tau_{\rm prec,p}$ for multiple values of $P_{\rm in}$. The timescale turnover point shifts from  $a_{\rm p} = 7$ AU for a 100 d binary to $a_{\rm p} = 3$ AU for a 5 d binary. The Kozai timescale on the planet does not change as the binary shrinks, because $P_{\rm p}$ and $P_{\rm out}$ are essentially static.

We simulated all stable systems from Sect.~\ref{sec:example_primordial} ($a_{\rm p}$ between 2 and 8 AU for $e_{\rm out} = 0$ and $a_{\rm p}$ between 4 and 7 AU for $e_{\rm out} = 0.25$) with shorter binary periods at 25, 10 and 5 d, and looked for ejected planets. The simulations were over 25 Myr, which covers possible Kozai cycles for the planet but not for the shrunken inner binary, for which the Kozai timescale becomes very long. This integration time is considered reasonable because planets were already seen to survive high eccentricity excursions by the wider primordial binary, which should have a more destabilising effect than shorter-period inner binaries.

The  three examples show:

\begin{itemize}
\item
For $a_{\rm p} = 4$ AU and $e_{\rm out} = 0$ the planet has a small eccentricity for $P_{\rm in}$ = 100, 25 and 10 d.  Around the primordial 100 d binary there is a bump in $e_{\rm p}$ near 10 and 20 Myr, corresponding to Kozai cycles of the inner binary. This bump is not seen for shorter-period binaries because the Kozai cycles are longer than 25 Myr. When the binary has shrunk to 5 d the tertiary star's strong influence on the planet, causes a significant raise in its eccentricity, yet it remains stable. There is an increase in the variation of $\Delta I_{\rm p,in}$ as $P_{\rm in}$ decreases.

\item
For $a_{\rm p} = 4$ AU and $e_{\rm out} = 0.25$ the results are qualitatively the same as in the case of a circular tertiary star.

\item
For $a_{\rm p} = 5$ AU and $e_{\rm out} = 0$ the planet obtains a high eccentricity when the binary has reached a 10 d period, but remains stable. For a 5 d period binary the planet does not survive for more than a couple of million years, because the very short-period inner binary is no-longer able to shield the planet from the perturbations from the tertiary star. This ejection is accompanied by a significant increase in $\Delta I_{\rm p,in}$
\end{itemize}

From the suite of simulations we derived rough stability limits as a function of $P_{\rm in}$. In Fig.~\ref{fig:stability_limits} we plot the inner and outer stability limits as functions of $P_{\rm in}$, for $e_{\rm out}$ = 0 (dark blue, solid) and 0.25 (light blue, dashed). Because the circumbinary precession timescale gets longer during the shrink, the tertiary star has an increasing influence over the planet. We found that this caused the outer stability limit to move inwards as $P_{\rm in}$ decreased. 

As $P_{\rm in}$ decreases the inner stability limit should also move inwards (e.g. $a_{\rm crit,CB} = 0.14$ AU for $P_{\rm in} = 5$ d from Eq.~\ref{eq:stability_limit_CB}). However, this is not a physically meaningful limit for any orbiting planets because they had to form around the wider primordial binary where $a_{\rm crit} = 2$ AU. Therefore, we define the inner stability limit as a constant, derived from the case of the primordial binary.

Overall, the shrinking of the binary via KCTF can be seen as a destabilising process, as only planets formed relatively close to the inner binary have a change of surviving the shrinking process. There is only a narrow 2 AU region where a planet could possibly form and survive, assuming that the tertiary star has a circular orbit. For $e_{\rm out} = 0.25$ the inner and outer stability limits are the same (at 4 AU) when $P_{\rm in}$ reaches 5 d, and hence the likelihood of both forming a planet in these circumstances and having it survive the KCTF process down to a 5 d inner binary is small.

%====================
%Subsection 
\section{Additional constraints on planet formation and evolution}
\label{sec:additional_effects}
\subsection{Protoplanetary disc environments}
\label{sec:discs}
%==================== 

Our analysis so far has been limited to n-body orbital dynamics. However planets are believed to form in discs and only under certain favourable conditions. These necessary conditions further restrict the possible range of disc radii that can allow planet formation in a stellar triple system.

A {\it circumbinary} disc is expected to have a truncated inner edge near the n-body stability limit \citep{artymowicz94}. However, it is considered theoretically challenging to form circumbinary planets close to this inner edge, due to secular forcing excreted by the binary that creates a hostile region for planetesimal accretion \citep{lines14}. The favoured theory is that planets are formed farther out in the disc in a more placid environment, before migrating inwards and halting near the inner truncation radius of the disc \citep{pierens13,kley14}. This theory naturally explains the pile-up of planets in Fig.~\ref{fig:stability_limit}. The fact that the abundance of circumbinary planets appears to be similar to that around single stars further supports this theory. A similar abundance implies a similar formation efficiency and environment, and a circumbinary disc only resembles a circumstellar disc far away from the inner binary. The implication is that a true inner limit for planet formation may be substantially  farther out than the limit of Eq.~\ref{eq:stability_limit_CB}. 

\begin{figure}  
\begin{center}  
\includegraphics[width=0.49\textwidth]{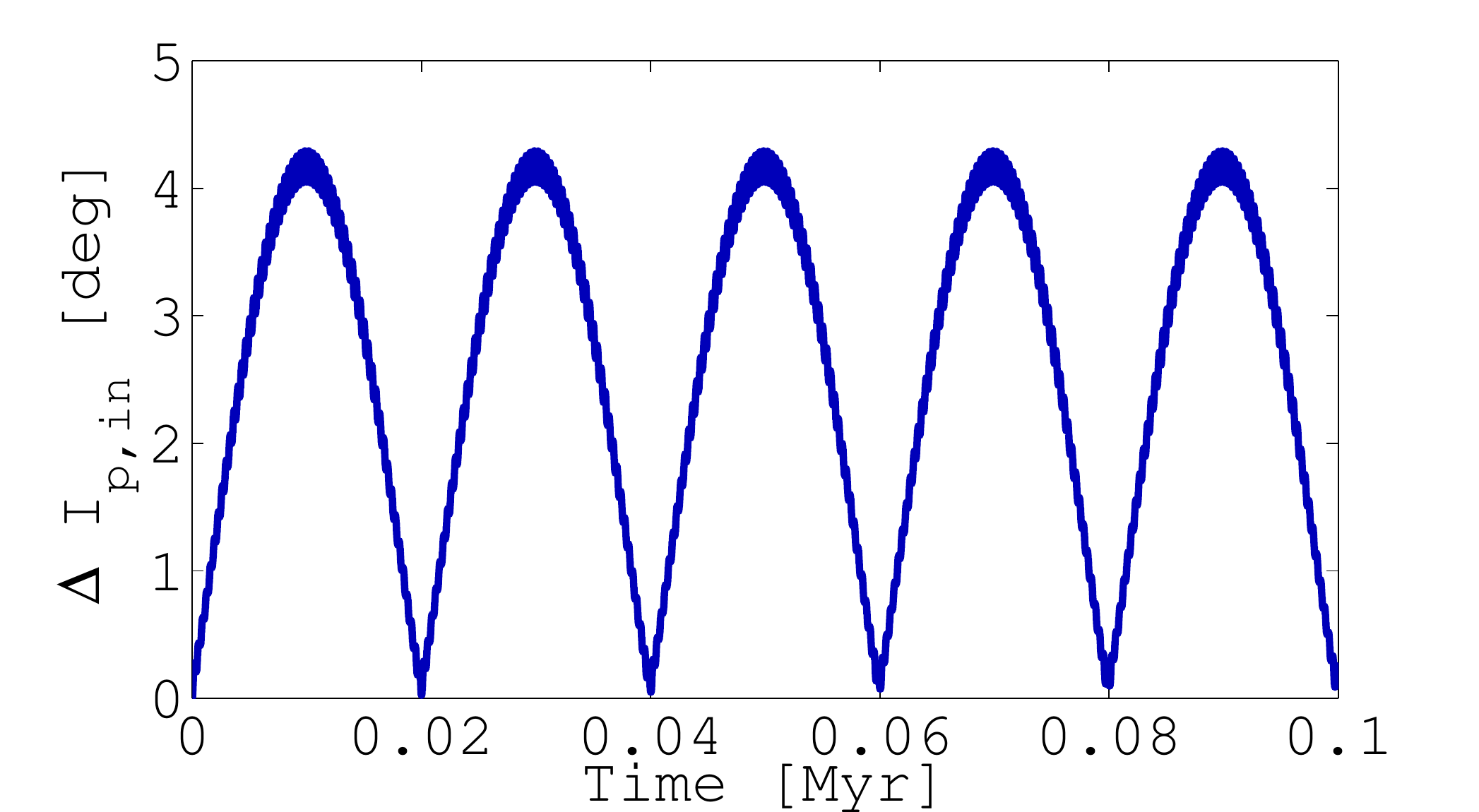} 
\caption{Short-term variation in $\Delta I_{\rm p,in}$ of a planet on an initially coplanar orbit at 7 AU around a primordial 100 d binary with a circular tertiary star at 100 AU, taken from the simulations in Fig.~\ref{fig:example_DeltaI}.}
\label{fig:zoomed_DeltaI}  
\end{center}  
\end{figure}

The formation of a {\it circumprimary} planet is also constrained by the interaction with the outer star, due to an outer truncation of the disc and perturbations on the planetesimals within. Like for the circumbinary case, the {\it outer} truncation radius in this case is also similar to the n-body stability limit \citep{artymowicz94}. However, here again the formation of planets in a circumprimary disc in the presence of a massive, inclined stellar companion can pose a significant theoretical challenge \citep{batygin11}. Theoretical studies expect the particles in the protoplanetary disc to undergo significant modulations in eccentricity and inclination with different periods and amplitudes at different radial distances, making planetesimal collisions destructive rather than accretive (e.g. \citealt{marzari09}). This matches our results which show stable orbits, but with large fluctuations in eccentricity and inclination. In Fig.~\ref{fig:zoomed_DeltaI} we demonstrate this further with a zoomed version of the variation of $\Delta I_{\rm p,in}$ from Fig.~\ref{fig:example_DeltaI} for $a_{\rm p} = 7$ AU and $e_{\rm out}$ = 0.25. The fluctuations are of a couple of degrees in amplitude and on a short timescale with respect to planet formation. It is possible, however, that if these fluctuations are sufficiently mild they may be damped by the self gravity of the disc \citep{batygin11} or viscous dissipation \citep{martin14}. 

Like in the circumbinary case, the perturbations on the disc will be strongest near the outer truncation radius. This suggests a more restrictive outer formation boundary. These considerations are in accordance with observations showing that the properties of planets in {\it circumprimary} orbits are roughly indistinguishable to planets around single stars only if the binary companion is farther than $\sim 100$ AU \citep{duchene09}, although the observational evidence for this is currently based on low number statistics. This finding is consistent with the scenario that giant planets are formed behind the snow line at at $\sim 5$ AU (e.g. \citealt{roberge10}), and that formation may be inhibited if the snow line is too close to or beyond the outer stability limit.

%Finally, the results of our simulations in Figs.~\ref{fig:example_eccentricity} and ~\ref{fig:example_DeltaI} show perturbations on planetary orbits on very short timescales. Similar perturbations imparted on a protoplanetary disc would likely lead to a chaotic environment for planet formation, due to high collisional velocities between planetesimals.

\subsection{Planet migration}
\label{sec:disc_migration}

Our analysis so far has only considered planets with static semi-major axes. However, as previously discussed, the favoured paradigm is that the circumbinary planets generally migrate inwards before being halted near the inner edge of the disc. The disc dispersal timescale \citep{alexander12} is expected to be much shorter than the KCTF timescale \citep{fabrycky07}, and hence any migration will only occur around the primordial binary.

This migration may have a positive or negative effect on planet survival, depending on the relative timescales of the inner binary Kozai cycles and disc migration. The disc truncation radius corresponds closely to the stability limit, which is in turn a function of $e_{\rm in}$. If the planet migrates quickly and reaches the inner edge of the disc while the inner binary is still circular, then it will have migrated in too far and will be ejected once $e_{\rm in}$ is subsequently excited during Kozai cycles. Alternatively, if the planet migrates slowly then the inner binary will have already undergone a full Kozai cycle and the disc will be truncated farther out, meaning that the planet will not get too close. According to Sect.~\ref{sec:shrinking_inner_binary}, having the planet near this edge will aid its later survival as the inner binary shrinks under KCTF.

\subsection{Multi-planetary systems}
\label{sec:multi_planet_systems}

Our analysis has assumed that there is only one circumbinary planet in the system, however with the discovery of Kepler-47 it is known that multi-planet circumbinary systems do exist in nature. Multi-planet stability has been analysed in circumbinary systems \citep{kratter13,hinse15} but never in the presence of a third star.

When there are multiple planets in the system then there may be planet scattering events (e.g. \citealt{chatterjee08} in the context of planets around single stars). This process may even be amplified in multi-stellar systems due to the eccentricity variations induced in the planets (like in Fig.~\ref{fig:example_eccentricity}), leading to more close encounters. Planet-planet scattering may lead to the orbits being pushed either inwards or outwards, potentially moving planets into unstable orbits. A planet may also induce an apsidal precession on other planets, which could possibly suppress Kozai perturbations from the tertiary star.

If KCTF proceeds unhindered by the multi-planet system, the host binary will shrink and the secular timescale it imparts on the planets will sweep over a large range. This evolution would be similar to a sweeping secular resonance \citep{nagasawa05}: when the difference in binary-forced precession frequencies of the planets matches the precession frequencies between the planets, it may lead to eccentricity excitation and perhaps destabilisation.

%==================	==
%section 
\section{Summary and discussion}
\label{sec:summary}
%====================

\subsection{The general argument}
\label{sec:general_argument}

We here summarise our explanation for the lack of observed transiting planets around very short-period binaries, due to the KCTF process. Below is a list of all the key ingredients and constraints.

\begin{enumerate}

\item The planet forms around a long-period primordial binary undergoing high eccentricity Kozai cycles induced by the distant companion.
\item To avoid ejection, the planet cannot be too close to the eccentric inner binary, so it can only reside on a relatively wide orbit: $a_{\rm p} \gtrsim 4a_{\rm bin}$.
\item However, the planet also cannot orbit too far from the inner binary, in order to have its Kozai cycles induced by the tertiary suppressed by the binary.
\item If the tertiary star is eccentric, as is often the case, then the region of stability is even smaller.
\item If the planet is sufficiently massive and on a sufficiently close but stable orbit, it may inhibit the Kozai modulation of the binary, and therefore prevent KCTF shrinkage.
\item Assuming that the binary orbit is able to shrink, some planets become unstable, because the suppression of the Kozai cycle by the binary (constraint (iii)) becomes too weak.  
\item Even if Kozai cycles of the planet are suppressed by the inner binary, it still obtains some variation in $\Delta I_{\rm p,in}$. 
\item Considering the disc environment within which planets form leads to further restrictions being imposed, because the inner and outer edges of the truncated disc are probably too chaotic for planet formation.
\item Finally, planet migration has the potential to either help or hinder the survival of planets.
\end{enumerate}

There is only a small parameter space where the planet can form, survive and not inhibit KCTF on the binary. In our example in this section only roughly Neptune-mass planets between 2 and 3 AU fulfilled all of the above criteria. Furthermore, the surviving planets were misaligned by over $10^{\circ}$. This triple star configuration would be even more restrictive if the perturbing strength of the tertiary were increased by (i) an increased mass ratio $M_3/(M_1+M_2)$, (ii) a decreased semi-major axis ratio $a_{\rm out}/a_{\rm in}$, (iii) an increased $e_{\rm out}$ or (iv) an increased mutual inclination $\Delta I_{\rm in,out}$, leading to higher eccentricity Kozai cycles. 

We conclude that most triple star systems evolving under KCTF are not conducive to hosting planets. Alternatively they host planets biased towards small masses, long periods and misaligned orbits, which are difficult to detect via transits.

The known circumbinary planets around wider binaries likely formed in a more placid environment, suggesting that companion stars either do not exist or are too far away to have an effect, like in PH-1/Kepler-64. We await future observations to confirm this. The fact that these planets have been found near the inner stability limit suggests that the binary orbit has not shrunk over a long timescale (e.g. via KCTF) after planet migration has finished.

Our analysis could be further extended by testing a wider range of orbital parameters to quantify different regimes of secular evolution of planets in triple star systems, similar to the approach taken in \citet{takeda08}, but this is beyond the scope of the current more qualitative investigation.

Whilst this general argument developed may account for the majority of the non-detections of transiting planets around short-period binaries, it may not be the only effect present. Very close binaries are tidally locked, which increases the rotation speed and can lead to increased stellar activity. The standard \kepler\ 30-minute cadence may lead to insufficient sampling at the shortest periods. And finally, \citet{martintriaud15} calculated that some of the closest eclipsing binaries may be sufficiently inclined with respect to our line of sight that transits by coplanar planets are not geometrically possible. 

%Note that stability tests throughout the paper were done on a timescale of millions of years. If additional instability were to develop on longer timescales, this would only strengthen our argument that the evolving triple star geometry is very restrictive.

%\subsection{A distribution of triple star systems}
%\label{sec:distribution}

%The analysis was extended by calculating the stability zone width for a distribution of triple star systems. We followed the method of \citet{fabrycky07} by constructing the initial orbits of the inner and outer binaries independently using the log-normal distribution of Duquennoy and Mayor 1991 (parameters). The mutual inclination between the two orbital planets was selected from an isotropic distribution (i.e. $\cos\Delta I_{\rm in, out}$ is uniform between -1 and 1). Our analysis was comparatively simpler than \citet{fabrycky07} because we used $M_1=M_2=M_3=1M_{\odot}$ and initially circular orbits for all stars, which is certainly an overly optimistic scenario for planet formation.

%====================
%Section 
\subsection{Looking ahead}
\label{sec:looking_ahead}
%====================

Some level of planet formation and survival might be possible in this shrinking binary scenario, however the bias towards planets with small masses and wide, misaligned orbits creates significant detection limitations. To find such planets it is likely that new techniques will need to be developed. The four years of \kepler\ photometric data are yet to be exhaustively searched, and may still yield discoveries of misaligned planets via eclipse timing variations \citep{borkovits11}, or transits on non-eclipsing binaries \citep{martintriaud14}, particularly on binaries found with the BEER technique \citep{faigler12}. A re-observation of the \kepler\ field by the future {\it PLATO} satellite may lead to transits on the existing sample of eclipsing binaries by new planets that have precessed into view. Radial velocity surveys can detect misaligned circumbinary planets, however short-period binaries are tidally locked, leading to an increased rotation velocity that decreases spectral precision.

There is also potential for gravitational lensing and direct imaging \citep{delorme13,thalmann14} discoveries of long-period circumbinary planets.  Finally, {\it GAIA} astrometry will be sensitive to potentially hundreds of giant circumbinary planets on periods of a few years, and can even provide a direct measurement of the mutual inclination \citep{salmann14}.

Continued observations, whether they reinforce this dearth or lead to surprising new discoveries, will allow this relatively new problem in exoplanetary astrophysics to shed new light on a fundamental field of stellar physics.

%====================
%Section 
\section*{Acknowledgements}
%====================

We appreciate the independent works of M\~{u}noz \& Lai and Hamer, Perets, \& Portegies Zwart. D.~V.~M. thanks the amazing ongoing support of Stephane Udry and Amaury Triaud. The authors are thankful for insightful conversations with Xavier Dumusque, Rosemary Mardling, Margaret Pan, Yanqin Wu and many others at the recent Triple Evolution \& Dynamics conference in Haifa, Israel. We also thank Tomer Holczer for his assistance in setting up the numerical computations. D.~V.~M. is funded by the Swiss National Science Foundation. The research leading to these results has received funding from the European Research Council under the EU's Seventh Framework Programme (FP7/(2007-2013)/ ERC Grant Agreement No.~291352), the ISRAEL SCIENCE FOUNDATION (grant No.~1423/11) and the Israeli Centers of Research Excellence (I-CORE, grant No.~1829/12). We made an extensive use of \href{http://adsabs.harvard.edu}{ADS}, \href{http://arxiv.org/archive/astro-ph}{arXiv} and thank the teams behind these sites.

\end{document}